\documentclass[12pt]{article}
\usepackage[utf8]{inputenc}
\usepackage{pifont}
\usepackage{fontawesome5}
\usepackage[T1]{fontenc}
\usepackage{amsmath,amssymb}
\usepackage{graphicx}
\usepackage{hyperref}
\usepackage{booktabs}
\usepackage{float}
\usepackage{placeins}
\usepackage{natbib}
\usepackage[margin=1in]{geometry}
\usepackage{epigraph}
\title{Asynchronous Barter, Periodic Reconciliation: Evidence for a \\
Scale-Free Commercial Network in the Indus Valley Civilization}

\author{
  Mahesh T C \\
  Independent Researcher \\
  ORCID: 0009-0002-8273-320X \\
  DOI: \href{https://doi.org/10.5281/zenodo.19394894}{10.5281/zenodo.19394894}
}

\date{\today}

\begin{document}

\maketitle

\epigraph{Yet to the steadfast eye of one standing upon the shore of things, looking
chiefly and most lovingly upon that side on which the sun shines and that we feel
joyously to be life, the heart is ever gladdened by the beauty, the exquisite
spontaneity, with which life seeks and takes on its forms in an accord perfectly
responsive to its needs. It seems ever as though the life and the form were absolutely
one and inseparable, so adequate is the sense of fulfilment.

Whether it be the sweeping eagle in his flight, or the open appleblossom, the
toiling work-horse, the blithe swan, the branching oak, the winding stream at its
base, the drifting clouds, over all the coursing sun, \textit{---form ever follows function---},
and this is the law. Where function does not change, form does not change. The
granite rocks, the everbrooding hills, remain for ages; the lightning lives, comes
into shape, and dies, in a twinkling.

It is the pervading law of all things organic and inorganic, of all things physical
and metaphysical, of all things human and all things superhuman, of all true
manifestations of the head, of the heart, of the soul, that the life is recognizable
in its expression, that form ever follows function.\textit{---This is the law---.}}
{Louis H. Sullivan, \textit{The Tall Office Building Artistically Considered}}

\begin{abstract}
The Harappan Civilization (c.\ 2600--1900 BCE) achieved extraordinary standardization
across a vast region without clear evidence of centralized rule. We propose that it was
a merchant commonwealth organized as a scale-free commercial network, with AP2PBS
(Asynchronous Peer-to-Peer Barter Settlement) and the Unicorn Logistics Confederation (ULC)
serving as inferential model labels for its coordinating institutions.

Four lines of evidence support this interpretation. (1) \textbf{Statistical}: Unicorn with
$\alpha \approx 1.80$, and offering stand frequencies follow a power-law distribution with
$\alpha \approx 2.3$--$2.6$, while seal counts across sites follow $\alpha \approx 1.55$;
all fit heavy-tailed organization better than exponential alternatives. (2) \textbf{Attribute
correlations}: non-random associations between offering stands and anatomical attributes,
including Cram{\'e}r's V patterns across market tiers, are consistent with deliberate
institutional design of visual confirmation bundles. (3) \textbf{Archaeological}:
seal evolution, contextual distributions, and warehouse logistics support the operational
requirements of long-distance asynchronous exchange. (4) \textbf{Architectural}:
Mohenjo-daro's halls, bath, warehouses, and cart-compatible gridiron are consistent with
infrastructure for commerce and coordination.

Taken together, these findings are consistent with AP2PBS and ULC as models of how
independent participants could standardize authentication, exchange value asynchronously,
and coordinate across distance without a hierarchical state. Scale-free topology also
offers a testable explanation for both Harappan longevity and its vulnerability to hub
failure and cascading decline. The argument is inferential: the archaeological record
does not directly identify AP2PBS or ULC by name, but its statistical, material, and
architectural patterns are compatible with this decentralized commercial model.
\end{abstract}

\section{Introduction}

The Harappan Civilization (also known as the Indus Valley Civilization) flourished
across a vast region of South Asia from approximately 2600 to 1900 BCE, spanning
over a million square kilometers across diverse ecological zones. The civilization
is renowned for its remarkable standardization: gridiron-planned cities, uniform
brick ratios, calibrated weights, and coordinated drainage systems. Yet the
archaeological record yields no palaces, no royal tombs, no temples, no monumental
inscriptions glorifying rulers, and no obvious centralized administrative authority.
This presents a fundamental paradox: How did a civilization encompassing such vast
territory achieve and maintain standardization without the hierarchical apparatus
(kings, priests, bureaucrats, armies) that characterizes other Bronze Age states?

Among the civilization's most distinctive artifacts are steatite seals, typically
measuring 2--3~cm per side, bearing finely carved iconographic motifs and an
undeciphered script. Over 3,700 seals are known from Harappan sites. Yet the
accompanying material record presents a striking contrast to other ancient
civilizations. Mesopotamian systems generated abundant clay sealings on packages
and administrative records---evidence of transaction documentation and centralized
archiving. By contrast, the Harappan record contains approximately 200 tablets and
a similar number of sealings. This scarcity is not a preservation accident: as
\citet{frenez2024} documents, the majority of known Harappan sealings come from a
single deposit at Lothal, preserved only due to accidental fire. Most sites yield
no sealings at all. This inverts the expected pattern: abundant seals (identity
tokens) but scarce sealings and tablets (transaction records). If Harappan seals
functioned like Mesopotamian seals (authenticating individual transactions), we
should expect abundant sealings corresponding to each transaction. Instead, we
find the opposite. This raises a fundamental question: What system would generate
abundant seals but systematically scarce (or absent except by accident) sealings and
tablets?

Additional puzzles compound this enigma. Monumental structures at Mohenjo-daro---the
Great Hall, Pillared Hall, and Great Bath---lack parallels to temples, palaces, or
administrative centers known from other civilizations. Their function remains unclear.
The most common seal motif is the unicorn, a wholly invented animal with no natural
prototype, appearing suddenly and dominating approximately 60\% of mature seals and
then disappearing as suddenly after the Mature Harappan period. The
offering stand---a non-animal element on the seal---exhibits far more variation than
any other attribute; it has been variously interpreted. \citet{kenoyer2005indus}
documents that "growth of cities such as Harappa appears to be directly linked to
increase in regional and long distance trade." He also notes that during the
Mature Harappan period ox carts were used in Harappan streets, where major streets
allowed two-way traffic while narrow gates were used at city gates, and that
"this feature appears to be distinct of Indus cities and has not been documented
in early urban centers of Mesopotamia or Egypt". Material culture evidence
shows toy carts and boats increase precisely when urban planning transitions from
organic to gridded layouts. Why would a logistics innovation (the cart) correlate
with urbanization? These patterns suggest an underlying organizing principle distinct
from state administrative models.

This paper investigates whether quantitative analysis of seal distributions and
archaeological evidence can clarify these puzzles. We employ network science,
statistical analysis of frequency distributions, and comparative archaeological
evidence to test whether the Harappan seal system encoded commercial organization
rather than reflecting primarily priest-, ruler-, or king-based administrative
models proposed in earlier interpretations.

Throughout this paper, the term ``guild'' is used heuristically to denote a
recurring commercial affiliation or trading entity represented within the
seal system. Given the archaeological contexts of seals and sealings—including
workshops, gateways, storage areas, jars, and packages—the term is used simply
as a convenient label for economically meaningful groups participating in exchange
networks.

\section{Assumptions, Frameworks and Mechanisms}

Our analysis is based on assumptions grounded in archaeological evidence
and employs two complementary theoretical frameworks. First, network externalities
explain how independent merchants converge on shared standards 
through self-reinforcing adoption without centralized enforcement. Second, power law 
mechanisms (Barab{\'a}si-Albert and Bentley-Hahn-Shennan models) characterize the 
statistical signatures that distinguish between preferential attachment (market-driven 
selection) and neutral drift (random cultural transmission). Together, these frameworks 
allow us to test whether Harappan seals exhibit patterns consistent with a merchant 
commonwealth organized by network effects, or alternative models of social organization.

\subsection{Core Assumptions}

This analysis rests on some assumptions grounded in archaeological 
evidence. First, the Harappan civilization operated as a merchant and trade network, 
not primarily as a state administrative hierarchy. As \citet{kenoyer2005indus} documents, 
``growth of cities such as Harappa appears to be directly linked to increase in 
regional and long distance trade.'' The material record confirms this: ox carts, 
gridded streets, and logistics infrastructure appear unique to Indus cities, and 
toy carts and boats proliferate precisely when urban planning transitions to gridiron 
layouts.

Second, seals functioned in commercial and logistic contexts, not in ceremonial 
or religious roles. \citet{mukhopadhyay2023} demonstrates that ``inscribed seals 
concentrate in functionally specific contexts,'' including city gates (traffic control), 
craft workshops (production control), and storage/warehouse systems (access control). 
Sealings appear ``attached to storage containers and warehouse locking systems,'' 
often alongside ``standardized Indus weights.'' \citet{frenez2024} documents 93 sealings 
from Lothal's warehouse excavation, with dense clusters reaching 70 sealings in a 
single square meter, confirming that seals authenticated goods in exchange contexts.

Third, seal distributions and attribute patterns reflect the underlying structure 
of the commercial network, not random artisanal variation or ceremonial convention. 
This assumption---that quantitative patterns in material culture encode organizational 
logic---follows network science approaches to archaeological data and is tested 
empirically against the data presented in this study.

And finally, we consider seals as a technology that enabled trade - initially it
was local, then regional and finally pan-Harappan.

Even though trade is considered diagnostic for Harappan civilization, how it was
organized remains under-theorized. We assume that there must have been a method
by which the USS functioned as a trade and logistics system for inter-regional trade.
It enabled merchants to ship goods securely across geographically dispersed, 
linguistically diverse networks, where packages could be verified at origin and 
claimed reliably at destination by trading partners with perhaps no shared language
or literacy. This contrasts with religious, administrative, socio-political or
ceremonial seal functions.

\subsection{Network Externalities}

The standardization patterns documented in this study suggest that the Sindhu
template---the gridiron-planned urban infrastructure, bathing and drainage systems,
standardized brick ratios, and calibrated weights system---and the unicorn
seal system exhibit classic network externalities, where the value of adopting
a standard increases with the number of other adopters. In the framework of
\citet{katz1985network}, once a critical mass of merchants adopted Harappan
weights and measures, defection became individually irrational regardless of
whether any central authority enforced compliance; the standard sustained
itself through switching costs and lock-in. Our proposed unicorn seal system
exhibits a complementary but distinct dynamic: the seal functions as a two-layer
signaling system, where the iconographic layer provides cross-linguistic routing
cues across what was likely a multilingual trade zone, while the script layer
encodes transaction-level markers between commercial principals. This composite
function---simultaneously a network recognition token and a compressed transaction
record---is consistent with what \citet{arthur1989competing} predicts as the
outcome of competing conventions under increasing returns. In that process, the
iconographic convention that maximized cross-network legibility accumulated
transactional volume, deepened counterparty recognition, and further entrenched
the convention, producing the preferential attachment dynamics documented in the
present study without requiring centralized enforcement.

\subsection{Power Law Mechanisms: Two Models}

Power law distributions are characterized by the probability density function:
\[
P(k) \propto k^{-\alpha}
\]
where $k$ is a quantity (e.g., frequency, size, degree) and $\alpha$ is the
power law exponent. This mathematical form produces the characteristic shape
of scale-free distributions: a few entities of very large size, followed by
a long tail of progressively smaller entities.

Power laws can arise from multiple distinct generative mechanisms. In cultural
and archaeological contexts, distinguishing between mechanisms provides insight
into the underlying social and economic dynamics. We focus on two mechanisms
particularly relevant to Harappan seals.

\subsubsection{Mechanism 1: Barab{\'a}si-Albert (BA) Preferential Attachment}

The Barab{\'a}si-Albert (BA) model \citep{barabasi1999emergence} describes growth
driven by preferential attachment: new nodes joining a network form connections
to existing nodes with probability proportional to the nodes' existing degree,
creating a self-reinforcing ``rich get richer'' dynamic. The pure BA model
produces a power law with exponent $\alpha = 3$. Empirical examples with confirmed
preferential attachment mechanism include scientific citation networks, internet
router topology, and actor collaboration networks \citep{jeong2003measuring}.
For Harappan seals, the BA model predicts that successful commercial entities
gain more links over time as new merchants preferentially adopt commercial
connections to established, reputable identities.

However, the Bianconi--Barab{\'a}si (BB) fitness model \citep{bianconi2001competition}
extends BA by allowing nodes to have intrinsic heterogeneous ``fitness'' values
that modulate their competitiveness in preferential attachment. Under BB dynamics,
a node with high fitness attracts links more readily than degree alone would predict,
while a node with low fitness attracts links more slowly. The BB model predicts
power law exponents in the range $2 < \alpha < 3$ depending on the distribution
of fitness values. For Harappan seals, BB is appropriate if commercial entities
possessed distinct reputational advantages (fitness) --- such as commodity
specialization, regional reach, or guild affiliation --- that influenced their
appeal to new traders independently of their current network position.

\subsubsection{Mechanism 2: Bentley-Hahn-Shennan (BHS) Neutral Drift / Random Copying}

The Bentley-Hahn-Shennan (BHS) model \citep{bentley2004random} describes cultural
transmission through unbiased (random) copying without selective preference. Under
pure neutrality, variants drift randomly in frequency, with exponent typically
$1.5 < \alpha < 2$. Empirical examples include baby names in U.S.
birth records \citep{bentley2004random}, pottery vessel forms among potters
without functional pressure, and artisanal variation where minor differences
accumulate through copying mistakes and individual preference.

A critical point: Bentley et al. explicitly identify that \textbf{non-random
correlations between attributes constitute deviation from neutrality}. They note
that while many pottery motifs followed neutral drift, ``specific pottery motifs,
perhaps representing prominent households, were not neutral.''

\section{Methods}

We fit power law models using the \texttt{powerlaw} Python package
\citep{alstott2014powerlaw}, which implements the framework of
\citet{clauset2009power}. Exponents ($\alpha$) were estimated by maximum
likelihood estimation (MLE), and the lower bound $x_{\min}$ was selected by
minimizing the Kolmogorov-Smirnov (KS) distance. We also compared power law
fits against alternative distributions (exponential, lognormal, stretched
exponential, and truncated power law) using log-likelihood ratio tests.

Archaeological samples are sparse: the number of distinct types per attribute,
or of sites and motifs, is typically in the tens, well short of the
$n > 100$ tail observations that \citet{clauset2009power} recommend for precise
parameter estimation, and more data cannot be recovered. We therefore do not
treat $\alpha$ as a precisely estimated constant. Three converging criteria are
used instead: (i) a bootstrap goodness-of-fit test, which asks whether the power
law is a plausible generative model for the observed distribution; (ii) a
linear trend on the log--log rank-frequency plot; and (iii) rejection of the
exponential, the natural null for a system without preferential
attachment, by its own bootstrap goodness-of-fit test as well as by likelihood
ratio. Where lognormal and truncated power law cannot be distinguished from the
power law, as expected with small samples, we say so. Results are reported as
consistent with scale-free structure, not as proof of a particular generative
mechanism.

To assess attribute co-occurrence, we computed Cram{\'e}r's V for associations
between offering stand type and each of the ten anatomical attributes (ear,
halter, neck, eye, head, hoof, tail, pizzle, leg, horn). Statistical
significance was evaluated with $\chi^2$ tests and Bonferroni correction for
multiple comparisons. Because these correlation tests require both a valid stand
value and a valid attribute value, they use pairwise deletion: seals coded as
`n/a' or `blank' for a given field are excluded only from that specific test.
Accordingly, the effective sample size is lower than the full 500-seal corpus
and varies by attribute.

\section{Harappan Seal Analysis}

We use the seal corpus catalogued by \citet{ameri2013regional}, which records
stamp seals from six sites: Mohenjo-daro, Harappa, Chanhu-daro, Lothal,
Kalibangan, and Banawali.  The corpus holds $N = 1{,}048$ motif-classified
seals in $n = 23$ iconographic motif classes.

\begin{figure}[H]
    \centering
    \includegraphics[width=0.8\textwidth]{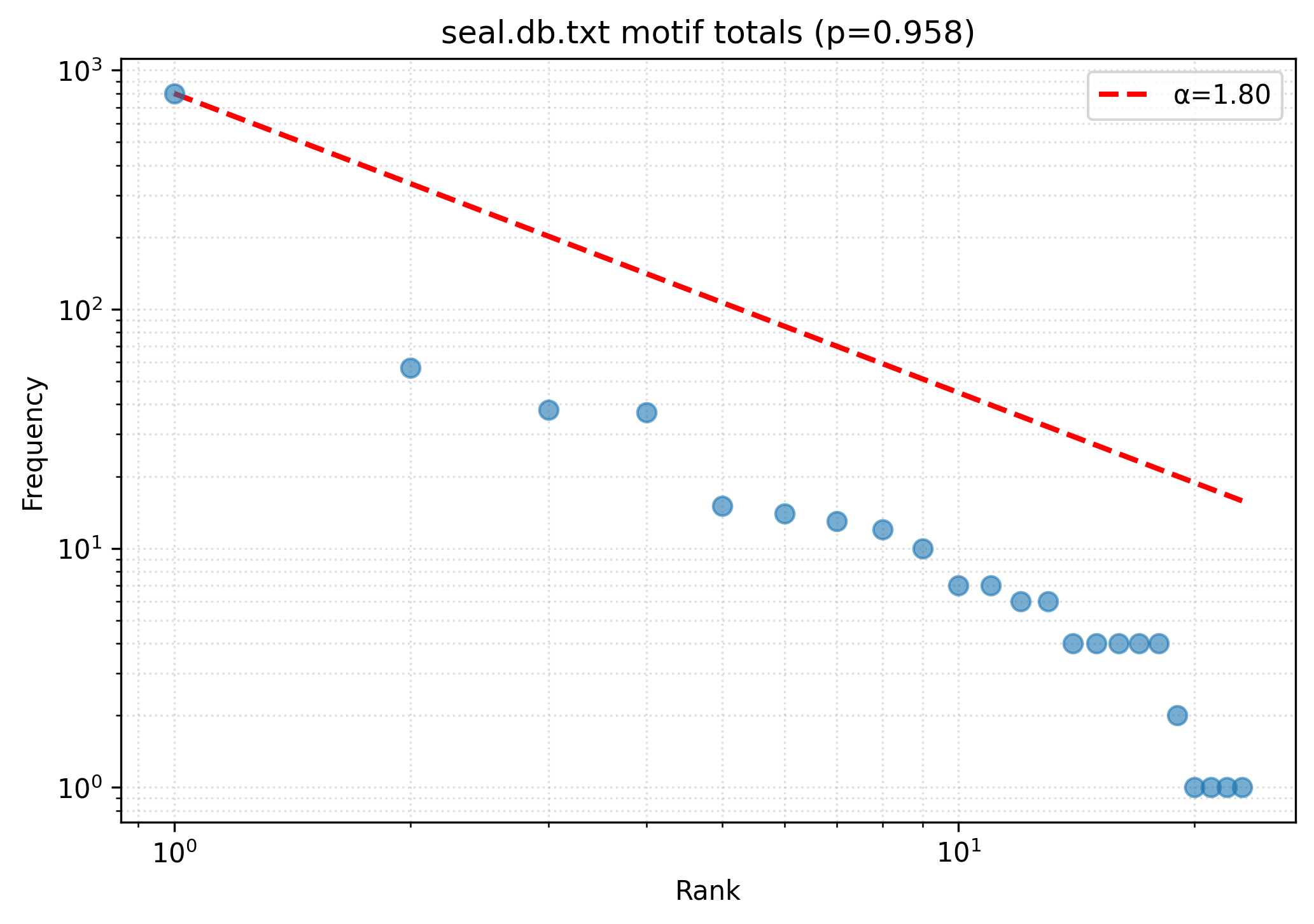}
    \caption{Log-log rank-frequency plot of the 23 iconographic motif classes (total seals per motif) with the fitted power law ($\alpha = 1.80$). The Unicorn motif dominates the head of the distribution.}
    \label{fig:seal_power_law}
\end{figure}

The distribution is highly skewed. The Unicorn alone accounts for $800$
seals ($\approx 76.3\%$); the other $22$ animal, human, and composite
motifs share the remaining $23.7\%$, most with single-digit counts.
We fit a power law ($\mathcal{P}(x) \propto x^{-\alpha}$) to the per-motif
totals by maximum likelihood, with $x_{\min}$ chosen by minimizing the
Kolmogorov--Smirnov (KS) distance \citep{clauset2009power,alstott2014powerlaw}.
Goodness of fit is assessed by a 2{,}500-draw parametric bootstrap. Results
appear in Tables~\ref{tab:powerlaw_stats} and
\ref{tab:model_comparison} and Figure~\ref{fig:seal_power_law}.

\begin{table}[htbp]
\centering
\caption{Power-Law Statistical Fit Parameters for Indus Seal Motifs.}
\label{tab:powerlaw_stats}
\begin{tabular}{ll}
\toprule
\textbf{Parameter} & \textbf{Value} \\
\midrule
Motif types & 23 \\
Total seals & 1048 \\
$\alpha$ & 1.7988 ($\pm 0.1883$) \\
$x_{\text{min}}$ & 4 \\
Motif types with $x \ge x_{\text{min}}$ & 18 \\
KS distance $D$ & 0.0678 \\
Goodness-of-fit $p$-value & 0.9584 \\
\bottomrule
\end{tabular}
\end{table}

\begin{table}[htbp]
\centering
\caption{Power law versus alternative distributions (loglikelihood ratio $R$; $R>0$ favors the power law), and the exponential goodness-of-fit bootstrap.}
\label{tab:model_comparison}
\begin{tabular}{lrr}
\toprule
\textbf{Alternative} & \textbf{$R$} & \textbf{$p$} \\
\midrule
Exponential & 22.8914 & 0.0050 \\
Lognormal & $-0.0070$ & 0.8421 \\
Truncated power law & $-0.0190$ & 0.8455 \\
\midrule
\multicolumn{3}{l}{Exponential GoF: $D = 0.5931$, $p < 0.0004$ (0/2500)} \\
\bottomrule
\end{tabular}
\end{table}

The fitted exponent $\alpha \approx 1.80$ ($1.5 < \alpha < 2.0$) lies in the
heavy-tailed, scale-free range. The power law is a plausible fit
($p = 0.958$) and decisively beats the exponential (Table~\ref{tab:model_comparison}),
which the exponential's own bootstrap also rejects ($p < 0.0004$). It cannot
be distinguished from the lognormal or truncated power law, as expected for
so small a sample.

In the Bianconi--Barab{\'a}si fitness model \cite{bianconi2001competition}, a node
whose intrinsic fitness far exceeds its competitors captures a macroscopic share
of all links. The Unicorn's share of over $76\%$ of seals, across geographically
disparate sites, is consistent with a single, highly fit standard outcompeting
regional alternatives.

Several caveats apply. Seal counts are subject to recovery and preservation
bias. Finally, small-type distributions in material culture can arise from
neutral drift \cite{bentley2004random}, but a single type holding a $76\%$ share
against 22 competitors deviates strongly from neutral expectation, pointing
to selection-driven, winner-take-all dynamics.

\section{Unicorn Seal System: Statistical Analysis}
\label{sec:unicorn_stats}

\subsection{Data}

We analyze published typological data from a sample of 500 unicorn seals across
19 archaeological sites, drawn from the comprehensive dissertation by
\citet{jamison2017}.  In a comprehensive diachronic study, \citet{jamison2017}
catalogued extensive stylistic variation in unicorn seals, decomposing the
iconography into eleven attributes---offering stand, ear, halter, neck, eye,
head, hoof, tail, pizzle, leg, and horn---each with multiple recorded types.

Table~\ref{tab:attributes} reproduces Jamison's numerical ranking of attributes by type count.

\begin{table}[h]
\centering
\caption{Numerical ranking of stylistic attributes by count of types
(after \citealt{jamison2017}, Table~7.3).}
\label{tab:attributes}
\begin{tabular}{@{}lcr@{}}
\toprule
Attribute & Rank & Number of Types \\
\midrule
Offering Stand & 1  & 114 \\
Ear            & 2  & 51 \\
Halter         & 3  & 49 \\
Neck           & 4  & 45 \\
Eye            & 5  & 42 \\
Head           & 6  & 39 \\
Hoof           & 7  & 36 \\
Tail           & 8  & 24 \\
Pizzle         & 9  & 19 \\
Leg            & 10 & 18 \\
Horn           & 11 & 5 \\
\bottomrule
\end{tabular}
\end{table}

Jamison's typological work is invaluable; it focuses on the organization
of seal production and its role in sociopolitical organization and stylistic
variations in them. We argue that Table~\ref{tab:attributes} itself reveals
a functionally significant distinction. The offering stand has \emph{more than
twice} as many types (114) as the next most variable attribute (Ear, 51). The
remaining attributes ---all anatomical parts of the unicorn animal---cluster in
a narrower range (5--51 types). This asymmetry is the key observation: a seal
reader at a warehouse does not inspect the ear, eye, or hoof independently of
the animal. One sees a unicorn. The anatomical variation reflects how different
artisans rendered the same animal, not distinct encoded information. The offering
stand, by contrast, is spatially separate from the unicorn and functionally
distinct---it is the one non-animal, non-script element on the seal, and it
is precisely the element with maximal variation.

We propose that this asymmetry reflects a functionally meaningful distinction
in the seal's information architecture. Examination of the frequency
distribution of offering stand styles reveals the signature of a power law
with a characteristic heavy tail: a small number of dominant stand
entity types appearing on many seals, and a long tail of rare types.

\subsection{Attribute Distributions}
\label{subsec:attributes}

We base our statistical analysis on the complete attribute data from
\citet{jamison2017}'s Appendix 7.1 rather than the binned summary tables
presented in his dissertation text. This distinction is methodologically
important: \citet{jamison2017} reports 114 offering stand types in his
dissertation narrative, whereas Appendix 7.1 records 139 offering stand
types. This discrepancy likely reflects a later aggregation step in the
narrative summary. For reproducibility, we treat Appendix 7.1 as the canonical
dataset because it preserves seal-level attribute observations, allowing direct
verification of type classifications and distribution fitting without potential
information loss from binning. The broader published corpus contains 500 seals,
but not every seal contributes equally to every downstream test: some records
lack usable stand or attribute values, and correlation analyses therefore work
with smaller effective $n$ values after pairwise deletion of `n/a' and `blank'
entries.

Table~\ref{tab:all_attributes} presents power law analysis results for all 11 seal attributes from Appendix 7.1.

\begin{table}[H]
\centering
\caption{Power law parameters for all 11 seal attributes. Attributes are ranked by exponent $\alpha$ (ascending). 
All goodness-of-fit tests performed using 2500 bootstrap iterations. 
$n_{\text{types}}$ = number of unique type classifications; 
$n_{\text{seals}}$ = count of seals with valid data for that attribute; 
$\alpha$ = power law exponent; 
$x_{\min}$ = optimal lower bound; 
$D$ = Kolmogorov-Smirnov distance; 
$p$-value = bootstrap goodness-of-fit $p$-value.}
\label{tab:all_attributes}
\begin{tabular}{@{}llrrrrrr@{}}
\toprule
Attribute & Types & Seals & $\alpha$ & $x_{\min}$ & $D$ & $p$-value & GoF \\
\midrule
Leg & 18 & 449 & 1.4175 & 1.0 & 0.0714 & 0.8808 & Yes \\
Tail & 25 & 351 & 1.6333 & 2.0 & 0.1125 & 0.5708 & Yes \\
Horn & 5 & 443 & 1.6407 & 9.0 & 0.1623 & 0.9028 & Yes \\
Halter & 55 & 470 & 1.7537 & 2.0 & 0.0801 & 0.7452 & Yes \\
Pizzle & 20 & 436 & 1.7606 & 5.0 & 0.1287 & 0.5200 & Yes \\
Head & 39 & 431 & 1.8454 & 4.0 & 0.1016 & 0.5040 & Yes \\
Stand & 139 & 413 & 2.3591 & 3.0 & 0.0514 & 0.9764 & Yes \\
Ear & 52 & 438 & 2.4210 & 7.0 & 0.1035 & 0.4064 & Yes \\
Neck & 72 & 458 & 2.5234 & 7.0 & 0.0925 & 0.4700 & Yes \\
Eye & 44 & 417 & 2.9340 & 13.0 & 0.1589 & 0.0372 & No \\
Hoof & 38 & 442 & 2.9396 & 13.0 & 0.0655 & 0.8936 & Yes \\
\bottomrule
\end{tabular}
\end{table}

\textit{Methodological note:} Horn exhibits an estimated $x_{\min}$ of 9.0
despite having only 5 total types, indicating that the power law fitting
algorithm may have encountered instability with extremely sparse type data.

The unicorn's anatomical attributes all show heavy-tailed frequency
distributions consistent with power-law or lognormal form, but their type counts
are too few to separate the two. This is a limit of the data, not a
finding of uniformity.

The critical observation is not distributional form per se but the magnitude
asymmetry. The offering stand, spatially separate from the unicorn on the seal
surface and not anatomically constrained by the logic of depicting a bovine,
generates 139 distinct types against a unicorn-attribute baseline clustering in
the 5--72 range. Even granting that anatomical attributes face physical constraints
on variation the stand's excess type diversity requires a generative explanation
beyond craft drift.

\subsubsection{Two Generative Mechanisms Operating on Seal Attributes}

The power law exponents and goodness-of-fit statistics reveal a fundamental
distinction in how different seal attributes are generated, with important
implications for understanding the information structure of the seal system.

\textbf{Neutral Cultural Drift (BHS mechanism, $\alpha \approx 1.5$--$2.0$,
\citet{bentley2004random}):} Five
attributes exhibit exponents characteristic of cultural drift: Leg
($\alpha = 1.42$), Tail ($\alpha = 1.63$), Halter ($\alpha = 1.75$),
Pizzle ($\alpha = 1.76$), and Head ($\alpha = 1.85$). These attributes
show the pattern predicted by \citet{bentley2004random} for neutral
cultural transmission: each generation of craftsmen departs slightly
from their training exemplar, producing heavy-tailed distributions in
which a small number of conventions dominate and a long tail of rare
variants accumulates. We note that the horn is diagnostic of a Unicorn
and it drifts to its canonical form.

\textbf{Preferential Attachment (BA mechanism, $\alpha > 2$,
\citet{barabasi1999emergence}):} Three anatomical attributes exhibit
exponents above the BHS range, consistent with preferential attachment
dynamics: Ear ($\alpha = 2.421$), Neck ($\alpha = 2.523$), and Hoof
($\alpha = 2.940$). All three pass goodness-of-fit tests (all $p > 0.40$).
Attribute correlation analysis in the following section explores whether
these high exponents reflect intentional bundling with stand identity.

\textbf{Attributes Excluded from Analysis:} Eye ($\alpha = 2.93$, $p = 0.0372$)
failed the goodness-of-fit test (p < 0.05), indicating its distribution does not
follow a power law. Horn was flagged earlier for methodological instability (5
types with estimated $x_{\min} = 9.0$). Both are excluded from further
interpretation.

\subsubsection{Offering Stand Distribution}
\label{subsubsec:offering_stands}

The offering stand is the most salient seal element that is spatially distinct
from the unicorn motif, and it dominates all other attributes in type diversity
(139 types) and in goodness-of-fit ($p = 0.9764$, the highest of the 11
attributes). Its exponent, $\alpha = 2.36$, lies in the preferential-attachment
range \citep{barabasi1999emergence}. Where nodes carry heterogeneous
fitness, as commercial guilds do in reputation, commodity specialization and
regional reach, the Bianconi--Barab{\'a}si model \citep{bianconi2001competition}
predicts $\alpha$ in the range $2$--$3$, consistent with the exponents observed
for identity-encoding attributes ($\alpha = 2.36$--$2.52$). Hubs of differing
fitness attract other nodes preferentially, so a small number of dominant guild
types account for most seals, and a long tail of rarer identities reflects
smaller or more specialised actors who link to the dominant hubs.

\begin{figure}[H]
    \centering
    \includegraphics[width=\textwidth]{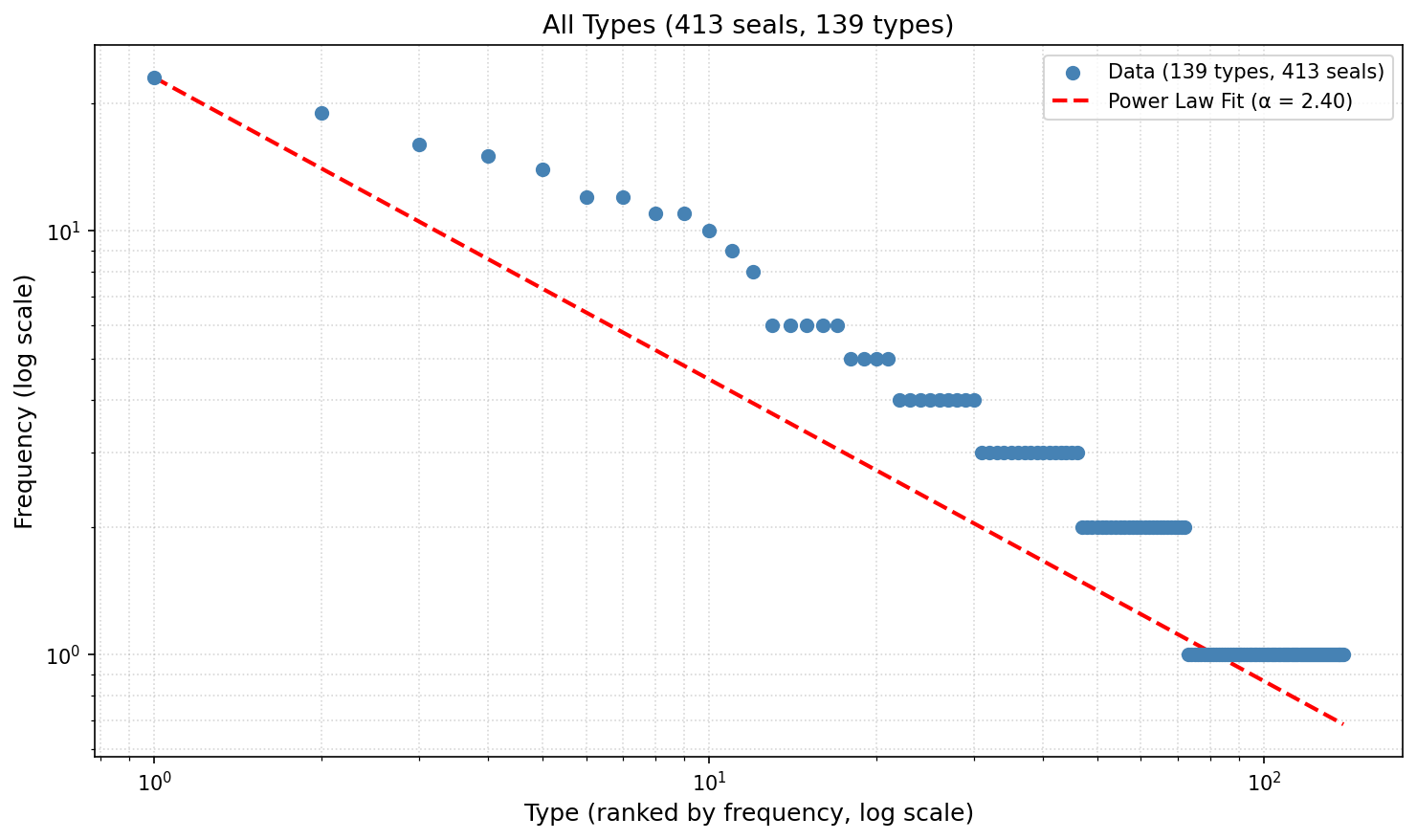}
    \caption{Log-log rank-frequency plot of all 139 offering stand types from the complete dataset \citep{jamison2017}.
The linear trend in log-log space is consistent with power-law structure despite the smaller tail sample ($n_{\text{tail}} = 46$).}
    \label{fig:appendix71}
\end{figure}

The bootstrap goodness-of-fit test passes the standard acceptance threshold
($p > 0.1$) of \citet{clauset2009power}, supporting power law as a plausible
generative model for the per-type distribution. The same test rejects the
exponential ($p_{\text{GoF}} < 0.001$, 0/2500). The lower bound $x_{\min} = 3.0$
leaves $n_{\text{tail}} = 46$ types in the power law regime, so
$\alpha = 2.3591$ is a preliminary estimate; the evidence rests on the
goodness-of-fit together with the linear log-log trend in
Figure~\ref{fig:appendix71}.

Table~\ref{tab:appendix_compare} summarizes likelihood ratio tests against four
alternative distributions. No alternative rules out the power law (minimum
$p = 0.108$, truncated power law), and it is favored over the exponential
($R = 1.25$, $p = 0.66$), though less decisively than for binned data because of
the smaller tail sample.

\begin{table}[H]
\centering
\caption{Distribution tests for offering stand data using complete Appendix 7.1 dataset.
First row shows exponential goodness-of-fit bootstrap test (rules out exponential).
Remaining rows show log-likelihood ratio tests: positive $R$ favors power law; negative $R$ favors alternative.}
\label{tab:appendix_compare}
\begin{tabular}{@{}lrrl@{}}
\toprule
Test & $R$ & $p$ & Interpretation \\
\midrule
Exponential (GoF bootstrap) & — & $<0.001$ & \textbf{Exponential rejected} \\
\midrule
Lognormal (vs PL) & $-0.90$ & 0.340 & No significant difference \\
Exponential (vs PL) & 1.25 & 0.657 & Power law favored, not significant \\
Stretched exponential (vs PL) & $-1.07$ & 0.319 & No significant difference \\
Truncated power law (vs PL) & $-1.29$ & 0.108 & No significant difference \\
\bottomrule
\end{tabular}
\end{table}

\subsection{Global Stand-Attribute Correlations}

We ask two questions of the Appendix 7.1 attribute data. First, does stand identity
predict the other Unicorn attributes across all usable seals? Second, does that
association hold across sites, or could it simply reflect local carving habits?
For the first question we compute the association between stand type and attribute
value using Cram{\'e}r's V (bias-corrected Bergsma-Wicher, 0--1 scale; near 0 means
independence). Permutation p-values with Bonferroni
correction ($\alpha = 0.005$) are used throughout. Because each test excludes only
seals with missing values for the relevant attribute, the effective sample size
varies across rows.

Table~\ref{tab:global_correlations} summarizes results for all 10 anatomical
attributes, with valid $n$ ranging from 379 to 407. Eight of ten attributes show
significant stand-dependent variation ($V = 0.30$--$0.48$), which is
strong for hand-carved, partly worn archaeological material. Leg and Halter
show the strongest associations ($V \approx 0.43$--$0.48$). Eye ($V = 0.22$,
$p = 0.066$) and Horn ($V = 0.29$, $p = 0.054$) do not reach corrected significance.
Horn has only five types and is nearly constant across seals, so it carries
little stand information; it is better read as a diagnostic feature of the Unicorn
itself than as part of a stand-specific bundle.

\begin{table}[H]
\centering
\caption{Global stand-attribute correlations using pairwise deletion.
$V$ = Cram{\'e}r's V (bias-corrected Bergsma-Wicher); $p_{\text{perm}}$ = permutation test p-value (5000 iterations);
$\alpha = 0.005$ with 10 attributes (Bonferroni correction).
Sample size varies by row because seals with `n/a' or `blank' entries are excluded only for the affected attribute.
Attributes ranked by $V$ (strongest associations first).
Status: \ding{51} = significant at corrected $\alpha$; \faExclamationTriangle{} = weak (one test only).}
\label{tab:global_correlations}
\begin{tabular}{@{}lcccl@{}}
\toprule
Attribute & $n$ & $V$ & $p_{\text{perm}}$ & Status \\
\midrule
LEG & 402 & 0.484 & $<0.001$ & \ding{51} \\
HALTER & 407 & 0.431 & $<0.001$ & \ding{51} \\
PIZZLE & 383 & 0.365 & 0.001 & \ding{51} \\
TAIL & 309 & 0.340 & 0.0038 & \ding{51} \\
HOOF & 393 & 0.337 & 0.0004 & \ding{51} \\
HEAD & 388 & 0.330 & 0.0008 & \ding{51} \\
EAR & 386 & 0.307 & 0.0002 & \ding{51} \\
NECK & 401 & 0.303 & 0.0008 & \ding{51} \\
HORN & 389 & 0.294 & 0.0538 & \faExclamationTriangle{} \\
EYE & 379 & 0.221 & 0.0656 & \faExclamationTriangle{} \\
\bottomrule
\end{tabular}
\end{table}

\subsubsection{Cross-Site Bundle Agreement}
\label{subsubsec:cross_site}

A pooled association can arise from local carving tradition alone: if carvers at
one site favor certain stand--attribute combinations and carvers at another site
favor different ones, stand and attribute will correlate in the pooled data without
any shared convention. The decisive test is whether the \emph{same stand} carries
the \emph{same attribute value} at \emph{different sites}.

For each attribute we take all pairs of seals that carry the same stand but come
from different sites, and record the fraction of pairs sharing the same attribute
value (the cross-site match rate). To judge this against chance, stand labels are
randomly shuffled among seals within each site. The shuffle preserves each site's
stand mix and attribute habits, so only a stand-linked pattern can exceed it. The
procedure uses a fixed random seed for reproducibility. Results appear in
Table~\ref{tab:cross_site}.

\begin{table}[H]
\centering
\caption{Cross-site bundle agreement. Match rate = fraction of same-stand,
different-site seal pairs sharing the same attribute value. Null = mean match
rate when stand labels are shuffled within sites. Different-stand = match rate for
different-stand, different-site pairs. $p$ is the permutation $p$-value
(smallest attainable value $\approx 0.0005$); Bonferroni $\alpha = 0.005$.
Attributes with few types (Horn) match often by chance.}
\label{tab:cross_site}
\begin{tabular}{@{}lrrrrrl@{}}
\toprule
Attribute & $n$ & Pairs & Same stand & Different stand & Null & $p$ \\
\midrule
HALTER & 407 & 550 & 0.287 & 0.127 & 0.135 & $<0.001$ \\
PIZZLE & 383 & 514 & 0.280 & 0.123 & 0.127 & $<0.001$ \\
HOOF   & 393 & 537 & 0.240 & 0.065 & 0.073 & $<0.001$ \\
TAIL   & 309 & 347 & 0.239 & 0.159 & 0.156 & 0.0025 \\
NECK   & 401 & 554 & 0.191 & 0.064 & 0.070 & $<0.001$ \\
LEG    & 402 & 550 & 0.513 & 0.325 & 0.346 & $<0.001$ \\
HEAD   & 388 & 538 & 0.152 & 0.076 & 0.080 & $<0.001$ \\
EAR    & 386 & 540 & 0.139 & 0.030 & 0.034 & $<0.001$ \\
EYE    & 379 & 473 & 0.116 & 0.048 & 0.049 & $<0.001$ \\
HORN   & 389 & 535 & 0.761 & 0.715 & 0.721 & 0.136 \\
\bottomrule
\end{tabular}
\end{table}

Nine of ten attributes show cross-site agreement beyond the within-site null,
all surviving Bonferroni correction. For Halter, the same stand shares a halter
type in 28.7\% of cross-site pairs against 13.5\% expected from local habit; for
Neck the figures are 19.1\% against 7.0\%, and for Hoof 24.0\% against 7.3\%. Horn
is the exception, consistent with its near-constant form. The same stand therefore
tends to carry the same attribute values at distant sites at two to three times the
chance rate, which local carving tradition alone does not explain.

Three limits apply. Match rates are well below 1, so this is a shared convention
with variation, not a fixed bundle per stand. Agreement is uneven across stands:
among widely attested stands some match closely on an attribute while others,
such as Stand \#44, show almost none, and per-stand samples (roughly 8--22 seals)
are too small for individual stand claims. Finally, a mobile, trained workshop
tradition would also produce this pattern, so the result establishes a shared
convention but does not by itself identify a coordinating institution; that
inference rests on the trade context and the model developed later.

\subsubsection{Tier-Specific Attribute Hierarchy}

We next analyze attribute-stand associations across three practical commercial
tiers: the top 5, 10, and 15 offering stand types. At each tier we compute
Cram{\'e}r's V between each attribute and the stand types present at that tier,
asking whether the association persists as the market widens. Here again, $n$ varies by attribute
because the tiered analyses use only seals assigned to the relevant stand tier
and then exclude `n/a' or `blank' values pairwise for each anatomical field.

\textbf{Methodological note:} The top-tier stands are predominantly
multi-site operations. Stands \#16, \#32, \#44, \#1, and \#30 each appear
across 3--5 archaeological sites spanning 1000+ kilometers, and pooled V
alone cannot separate a shared convention from local workshop tradition (see the
cross-site test above). Because both $n$ and the number of
stand types change from tier to tier, V values are not strictly comparable across
tiers, and the tier trends are descriptive. If V \emph{decreases} with tier size,
core attributes stay coupled to stand identity while attribute diversity grows with
market expansion.

Table~\ref{tab:attribute_hierarchy} summarizes Cram{\'e}r's V for all 10 attributes
across the three tiers. All associations are significant at Bonferroni-corrected
$\alpha = 0.05/10 = 0.005$.

\begin{table}[H]
\centering
\caption{Attribute-stand association strength (Cram{\'e}r's V) across three market tiers.
$n$ = number of seals; $k$ = number of stand types present at each tier.
All associations significant at Bonferroni-corrected $\alpha = 0.005$.
Attributes ranked by mean $V$ across all three tiers (strongest associations first).
Decreasing $V$ across tiers suggests core attributes remain
tightly coupled despite increasing stand-type diversity.}
\label{tab:attribute_hierarchy}
\begin{tabular}{@{}lcccccc@{}}
\toprule
Attribute & \multicolumn{2}{c}{Top 5} & \multicolumn{2}{c}{Top 10} & \multicolumn{2}{c}{Top 15} \\
\cmidrule(lr){2-3} \cmidrule(lr){4-5} \cmidrule(lr){6-7}
 & $n=86$ & $V$ & $n=140$ & $V$ & $n=171$ & $V$ \\
\midrule
NECK & 86 & 0.524 & 140 & 0.432 & 171 & 0.354 \\
HALTER & 86 & 0.492 & 139 & 0.389 & 171 & 0.395 \\
HORN & 85 & 0.332 & 138 & 0.371 & 168 & 0.464 \\
HOOF & 82 & 0.470 & 135 & 0.397 & 165 & 0.296 \\
EAR & 85 & 0.418 & 137 & 0.348 & 167 & 0.321 \\
EYE & 82 & 0.385 & 131 & 0.378 & 160 & 0.286 \\
LEG & 86 & 0.407 & 140 & 0.337 & 171 & 0.290 \\
PIZZLE & 84 & 0.397 & 136 & 0.302 & 165 & 0.223 \\
HEAD & 85 & 0.332 & 137 & 0.338 & 168 & 0.251 \\
TAIL & 65 & 0.374 & 108 & 0.290 & 135 & 0.218 \\
\bottomrule
\end{tabular}
\end{table}

The tiered pattern is descriptive. In the top five tier (stands with 23, 19, 16, 15
and 14 seals, together 87 seals or about 21\% of the corpus), Neck ($V = 0.524$) and
Halter ($V = 0.492$) are the most strongly stand-linked attributes, with Hoof
third ($V = 0.470$). Neck then weakens as the tier widens (0.354 in the top 15),
and Hoof falls more sharply (0.296), whereas Halter stays comparatively stable
(0.395). Globally Leg and Halter lead ($V = 0.484$ and $0.431$) and Neck ranks
eighth ($V = 0.303$), so Neck's association with stand is concentrated in the hub
tier. One reading is that the hub-tier bundle did not extend uniformly to rarer
stands while Halter linked stand identity across the market; the tier comparison
cannot establish this, since $n$ and the number of stands differ by tier.

The pooled V results and the cross-site test (Table~\ref{tab:cross_site}) together
indicate a shared convention for pairing stand and attributes: eight of ten
attributes associate with stand identity, and nine of ten agree across sites
beyond the within-site null. The agreement is partial, so we describe a convention
with variation, not a fixed bundle per stand. Whether a coordinating institution
enforced it is an inference from the trade context and the model developed below,
not a result of these statistics alone. We agree with \citet{jamison2017}, who
argues that "unicorn seal production was decentralized, it is important to
reiterate what is actually meant by this statement. It does not imply that
unicorn seal production and use were not regulated and controlled."

\FloatBarrier

\subsection{Diachronic Statistics}

To examine whether the power-law structure is stable across the Harappan
chronology, we assigned each seal in Appendix 7.1 a period using Tables 9.3, 9.5
and 9.6 of \citet{jamison2017} (merged file \texttt{appendix7.1.diachronic.txt}).
Most seals were recovered before modern stratigraphic controls, so
\citet{jamison2017} assigned periods by comparing seal and inscription carving
styles with stratigraphically controlled HARP material from Harappa, the most
reliable dating currently available for the corpus. The bootstrap goodness-of-fit
test uses 1000 iterations with random seed 42. For period 3A (3 stand types,
4 seals), automatic $x_{\min}$ selection in \texttt{powerlaw} is unstable, so a
fixed $x_{\min}=1$ is used.

\begin{table}[H]
\centering
\caption{Diachronic power law analysis of offering stand distributions by period.
$D$ = Kolmogorov-Smirnov distance; $p$ = bootstrap goodness-of-fit $p$-value
(1000 iterations, seed = 42). $^*$Period 3A has only 4 seals and is not
informative.}
\label{tab:diachronic_power_law}
\begin{tabular}{@{}lrrrrrr@{}}
\toprule
Period & Stands & Seals & $\alpha$ & $D$ & $x_{\min}$ & $p$-value \\
\midrule
3A (2600--2450 BCE)$^*$ & 3 & 4 & 2.6898 & 0.1163 & 1.0 & 0.5250 \\
3B (2450--2200 BCE) & 27 & 45 & 2.2649 & 0.1333 & 1.0 & 0.0160 \\
3C (2200--1900 BCE) & 42 & 93 & 2.0906 & 0.0665 & 1.0 & 0.2140 \\
\bottomrule
\end{tabular}
\end{table}

\begin{figure}[htbp]
    \centering
    \includegraphics[width=\textwidth]{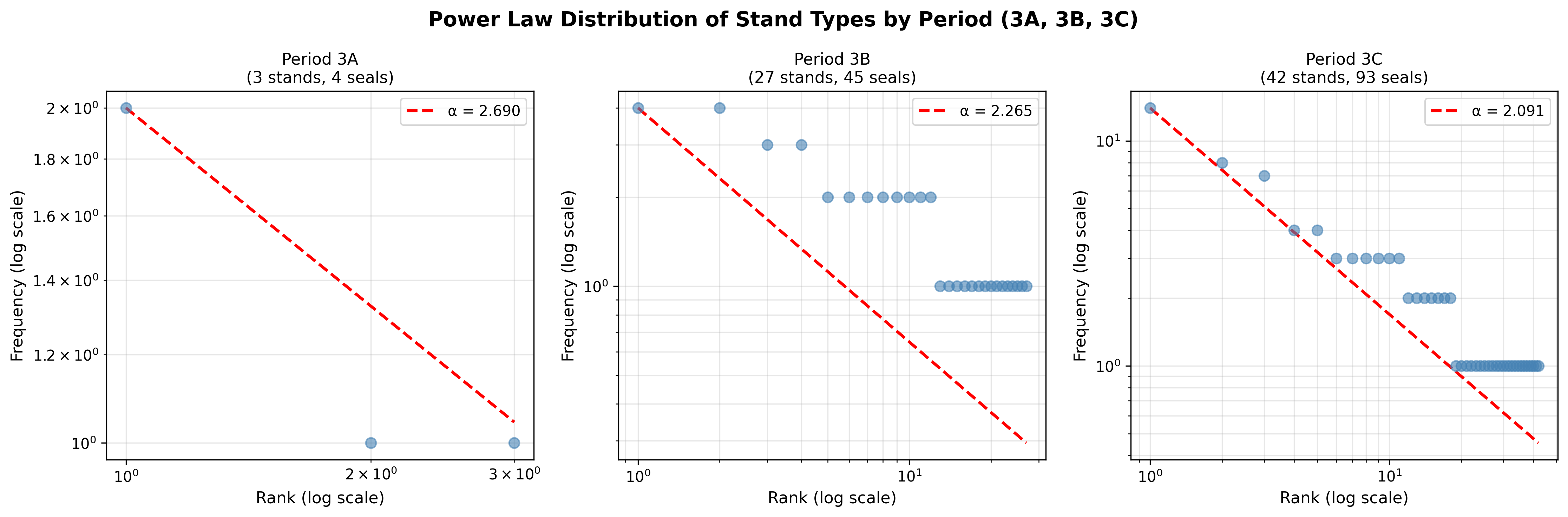}
    \caption{Diachronic rank-frequency distributions of offering stands across the
    three Harappan periods. The later phase is the most compatible with a
    heavy-tailed distribution; the earlier phases are limited by sparse data or
    weaker goodness-of-fit.}
    \label{fig:diachronic_power_law}
\end{figure}

Estimated exponents exceed 2 in all three periods, but goodness-of-fit is uneven
(Table~\ref{tab:diachronic_power_law}, Figure~\ref{fig:diachronic_power_law}).
Period 3C is compatible with a power law ($p = 0.214$), period 3B fails the
bootstrap test ($p = 0.016$), and 3A is uninformative. The diachronic samples are
small, so these results are suggestive and additional stratified seals would allow
a sharper test of when power-law structure emerged. Stand diversity rises from 3 to
27 to 42 types across the periods, which is consistent with expanding trade
networks and growing commercial specialization, and the heavy-tailed form
in the later phase suggests the stand system was not a late or ephemeral feature.

\FloatBarrier

\subsubsection{Diachronic Trajectory of the Unicorn Seal System (USS)}
\label{subsubsec:diachronic_scurve}

Seal counts and stand variety across the three periods are tabulated below.

\medskip
\begin{center}
\begin{tabular}{|l|c|c|c|}
\hline
 & \textbf{3A} & \textbf{3B} & \textbf{3C} \\
\hline
USS seal count & 4 & 45 (+1025\%) & 93 (+107\%) \\
\hline
Stand variety (types) & 3 & 27 (+800\%) & 42 (+56\%) \\
\hline
\end{tabular}
\end{center}
\medskip

The trajectory has the shape of an S-curve: a lag phase in 3A, rapid expansion
in 3B, and slower growth in 3C. This is consistent with network-adoption dynamics
in which new participants joined a standard whose value rose with its use
(preferential attachment), and with a slowing of new stand types in 3C as the
roster of commercial identities was largely covered, while seal production
continued to grow. We treat this as a suggestive heuristic and not a fitted model.
Three points cannot support a logistic regression, the counts come from a sample
of dated seals and not from the whole system, and a larger 3C sample can by itself
produce more stand types. More dated seals are needed to tell a robust growth
curve from a sparse-sample artifact.

\section{Site Distribution: Statistical Analysis}
\label{sec:site_stats}

\subsection{Data}

Table~\ref{tab:sites} shows the distribution of analyzed unicorn seals across the 19 sites in the sample.

\begin{table}[H]
\centering
\caption{Distribution of unicorn seals in the sample by site and state of preservation
(reproduced from \citealt{jamison2017}, Table~7.1).}
\label{tab:sites}
\begin{tabular}{@{}lrrr@{}}
\toprule
Site & Seals analyzed & Number of seals published & Number of intact seals \\
\midrule
Harappa       & 193 & 312 & 110 \\
Mohenjo-daro  & 191 & 832 & 408 \\
Lothal        & 35  & 45  & 19 \\
Kalibangan    & 24  & 26  & 13 \\
Chanhu-daro   & 16  & 21  & 11 \\
Bagasra       & 7   & 7   & 5 \\
Dholavira     & 6   & 6   & 6 \\
Allahdino     & 5   & 5   & 4 \\
Balakot       & 4   & 4   & 1 \\
Nausharo      & 4   & 4   & 4 \\
Rakhigarhi    & 3   & 3   & 2 \\
Banawali      & 3   & 3   & 2 \\
Farmana       & 2   & 2   & 2 \\
Nindowari     & 2   & 2   & 2 \\
Jhukar        & 1   & 1   & 1 \\
Kot Diji      & 1   & 1   & 0 \\
Lohumjo-daro  & 1   & 1   & 0 \\
Pabumath      & 1   & 1   & 0 \\
Surkotada     & 1   & 1   & 1 \\
\midrule
\textbf{Totals} & \textbf{500} & \textbf{1277} & \textbf{592} \\
\bottomrule
\end{tabular}
\end{table}

\subsection{Site Seal Distribution}

\subsubsection{Geographic Distribution of Seals}

The distribution of seal counts across archaeological sites follows a power law:

\begin{itemize}
    \item Power law exponent: $\alpha \approx 1.55$
    \item Lower bound: $x_{\min} = 2$ (5 single-seal sites excluded)
    \item KS distance: $D = 0.094$
    \item Goodness-of-fit $p = 0.76$ (bootstrap, 2500 iterations)
\end{itemize}

The fit is strong ($D = 0.094$, $p_{\text{GoF}} = 0.76$) and outperforms
an exponential distribution ($R = 9.09$, $p = 0.019$). The high goodness-of-fit $p$-value
means the power law cannot be rejected as a plausible generative model for the data
\citep{clauset2009power}. The optimal lower bound is $x_{\min} = 2$; the power law holds across
fourteen sites with 425 seals (five single-seal sites excluded).

Table~\ref{tab:site_compare} summarizes the likelihood ratio tests against four alternative
distributions.

\begin{table}[h]
\centering
\caption{Distribution comparison tests for site seal counts.
Positive $R$ favors power law; negative $R$ favors the alternative.}
\label{tab:site_compare}
\begin{tabular}{@{}lrrl@{}}
\toprule
Alternative & $R$ & $p$ & Interpretation \\
\midrule
Lognormal & $-0.35$ & 0.520 & No significant difference \\
Exponential & 9.09 & 0.019 & Power law significantly better \\
Stretched exponential & $-0.39$ & 0.501 & No significant difference \\
Truncated power law & $-0.61$ & 0.271 & No significant difference \\
\bottomrule
\end{tabular}
\end{table}

\begin{figure}[H]
    \centering
    \includegraphics[width=\textwidth]{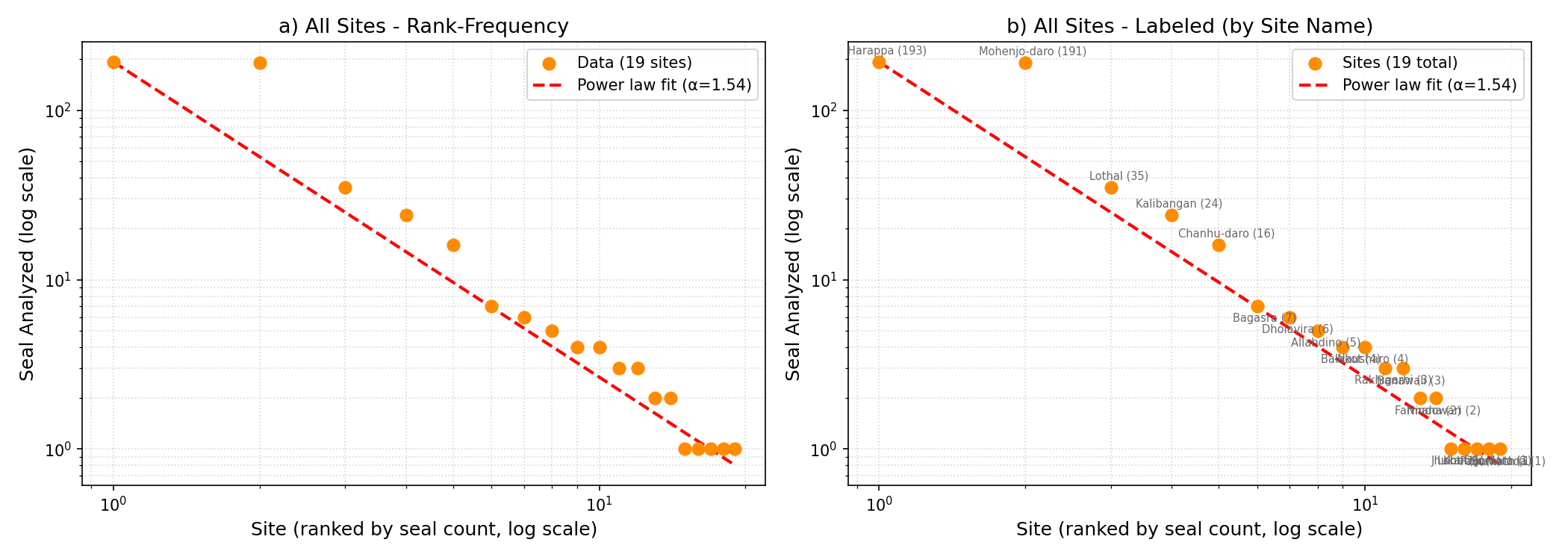}
    \caption{Log-log rank-frequency plot of seal counts by archaeological site.
    (a)~All 19 sites. (b)~Sites labeled with names and seal counts. Harappa and
    Mohenjo-daro emerge as clear hub sites.}
    \label{fig:sites}
\end{figure}

The two most extensively excavated sites---Harappa (193 seals) and Mohenjo-daro
(191 seals)---together account for 76.8\% of analyzed seals from only 10.5\% of sites.
This concentration may partly reflect excavation history; sites such as Rakhigarhi,
potentially the largest Harappan settlement by area, could alter the distribution as
new seal data are published. The lower exponent ($\alpha \approx 1.55$)
compared to the commercial entity distribution reflects a heavier tail and even
greater concentration at the site level: the top few cities are disproportionately
dominant trade hubs.

\begin{figure}[H]
    \centering
    \includegraphics[width=\textwidth]{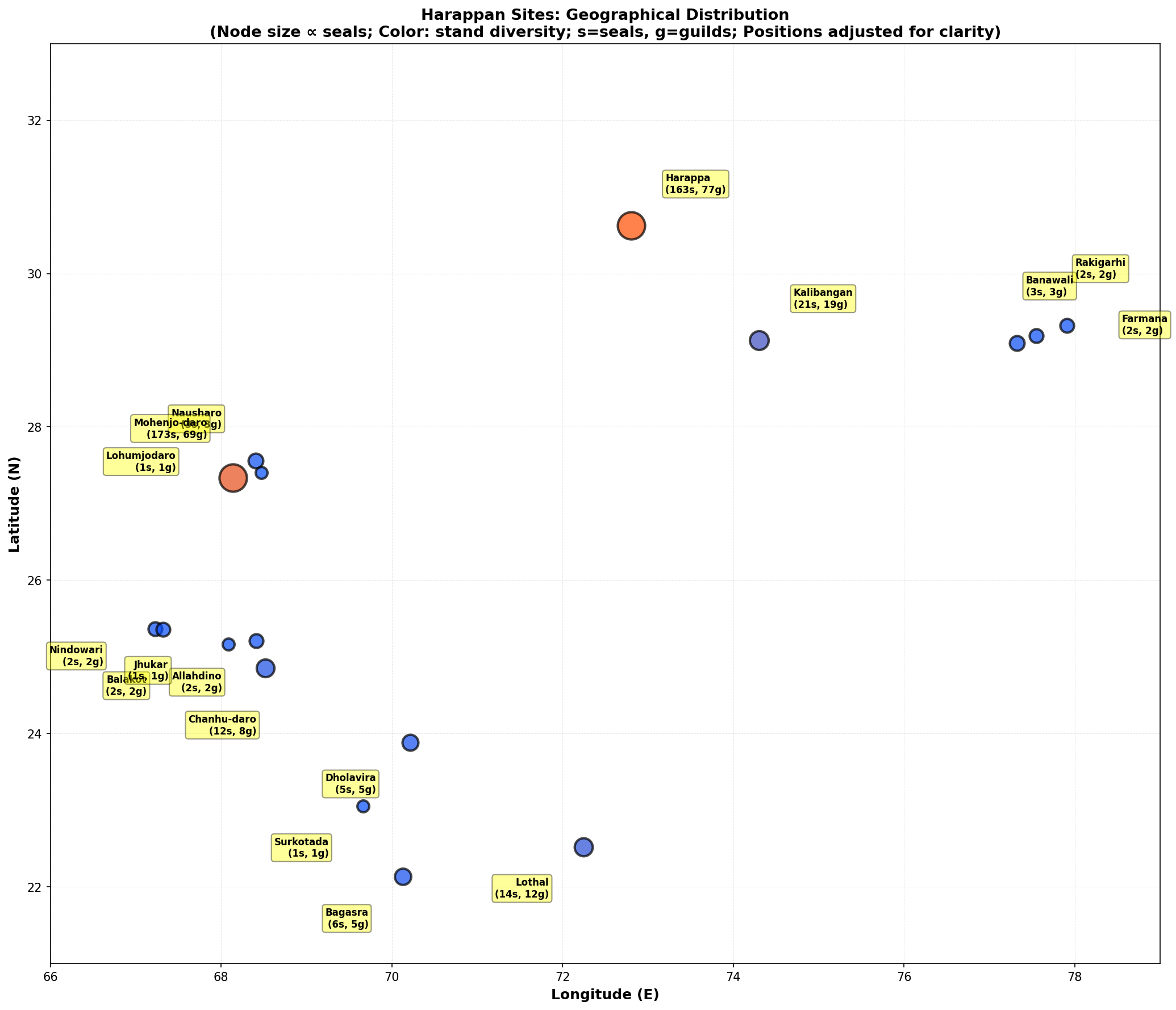}
    \caption{Geographic distribution of seal types across archaeological sites.}
    \label{fig:geographical_network}
\end{figure}

\subsubsection{Multi-Site Hub Guilds and Network Topology}

\begin{figure}[H]
    \centering
    \includegraphics[width=\textwidth]{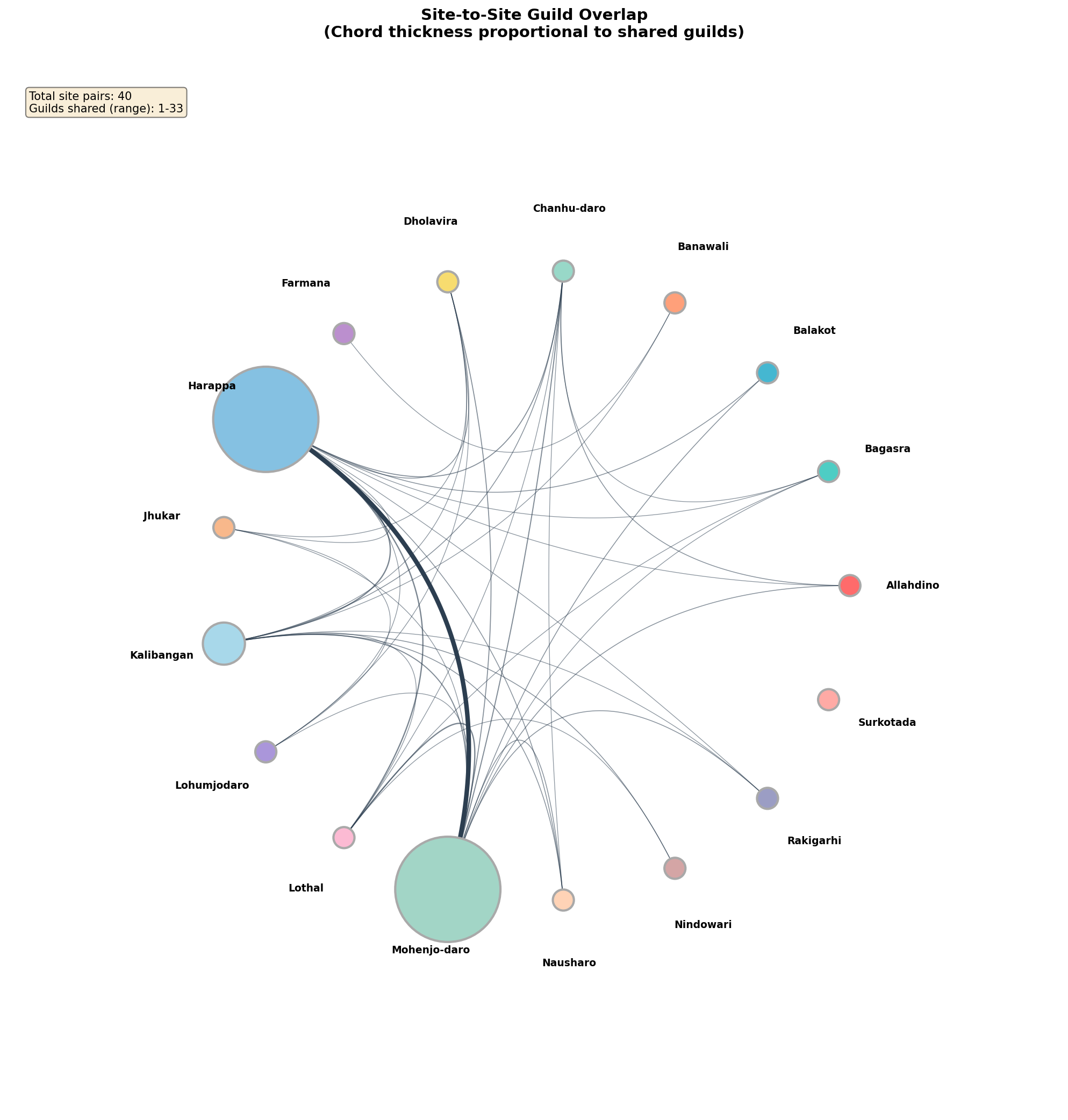}
    \caption{Offering-stand overlap between archaeological sites. Chord thickness
    represents the number of shared stand types; node size represents each site's
    seal count.}
    \label{fig:site_overlap_chord}
\end{figure}

Of the 139 stand types, 64\% are recorded at one site and 36\% at two or more.
Stands \#25 and \#32 each occur at five geographically dispersed sites, while
Harappa and Mohenjo-daro share the greatest overlap in stand types
(Figure~\ref{fig:site_overlap_chord}). These distributions indicate unequal
geographic reach among stand types, although they cannot by themselves establish
branch offices or direct trading relationships.

\subsubsection{Singleton Guild Distribution}

\begin{figure}[H]
\centering
\includegraphics[width=0.9\textwidth]{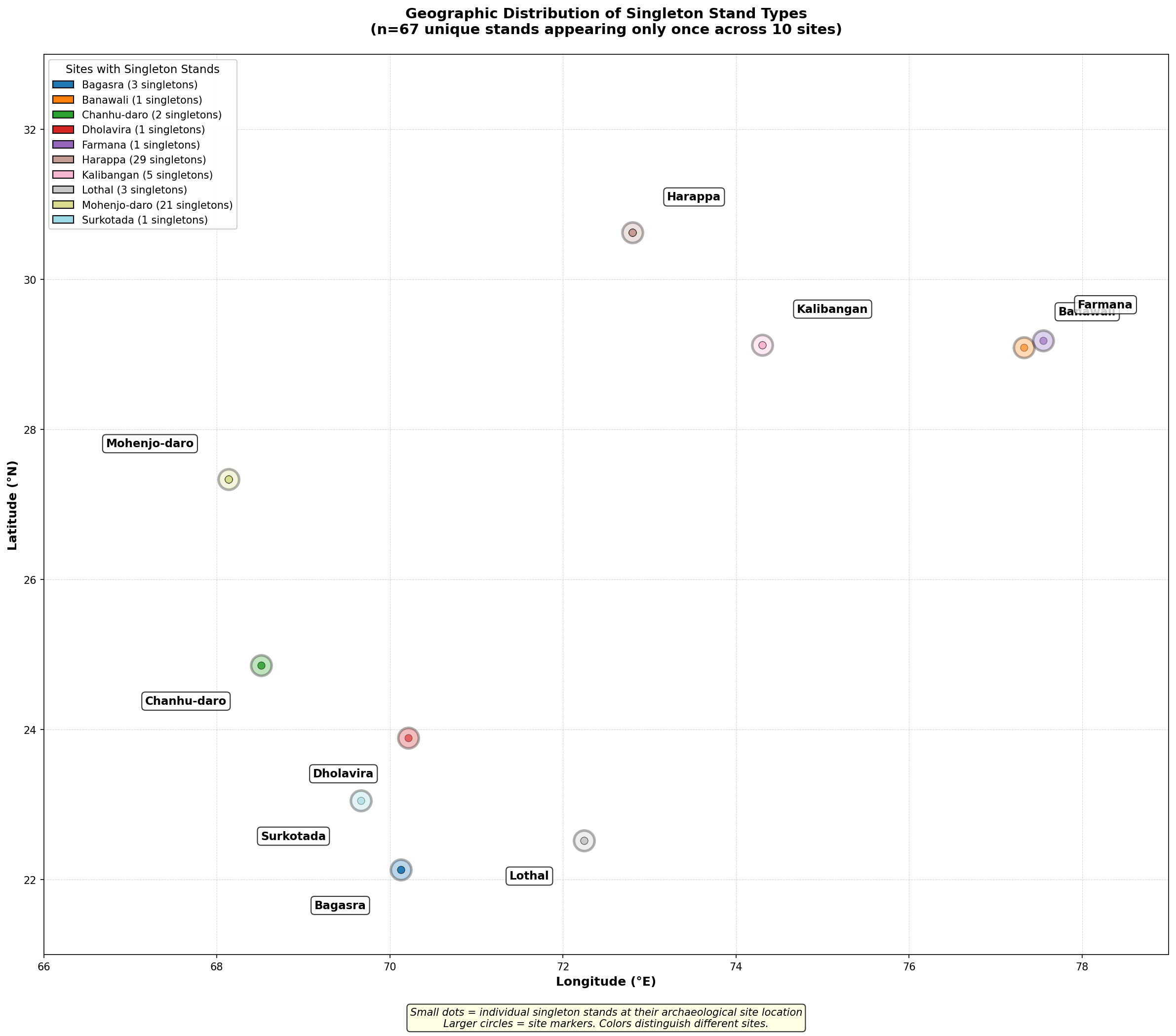}
\caption{Geographic distribution of singleton stands across Harappan sites.
Singletons (stand types appearing at only one site) are widely dispersed across
the network rather than concentrated at hub locations.}
\label{fig:singleton_distribution}
\end{figure}

The geographic distribution of singleton stands---stand types appearing at
only a single site---reveals a complementary pattern that illuminates the
structure of Harappan commerce. Figure~\ref{fig:singleton_distribution} maps the
spatial distribution of these 50+ unique stand types, showing that singleton
stands are not concentrated at a single hub but instead are distributed across
the entire geographic network, including peripheral and satellite sites.

\section{Archaeological grounding of the statistical analysis}

The seal system of the Indus civilization evolved over nearly a thousand years as
increasing trade generated problems related to scaling in long-distance commerce.
This evolution was not just decorative drift---it was iterative engineering that
extended the reach and reliability of trade networks across unprecedented distances.

\subsection{Seal Evolution as Trade Scaling Solution}

Archaeological work by \citet{kenoyer2010inscribed} and colleagues, based on
excavations at Harappa and other sites, documents three distinct phases of
seal development. We argue that these phases correspond to changing trade
network demands.

\subsubsection{Geometric Seals (Early Harappan, c.\ 3200--2600 BCE): Local Trade among Known Parties}

The earliest Indus seals bore geometric designs—intersecting circles,
cross-hatching, concentric squares, swastika and similar motifs. These seals
appear in context with clay sealings at sites such as Harappa (Kot Diji phase),
demonstrating use in verification and authentication from the earliest pre-urban
phases.  Geometric seals perhaps functioned effectively for small, local networks
where parties knew each other personally and relationships could possibly be
family-based. In these networks, agreements were verbal and did not require
written documentation. The seal indicated ownership and distinguished goods
from similar consignments, and the sealing (clay impression with cord marks)
provided tamper-proof evidence of package integrity during short-distance
movement as well as record-keeping.

Material evidence reflects this early phase: geometric seals were manufactured from
diverse materials—bone, terracotta, steatite, and faience—suggesting experimentation.
This trial-and-error approach gradually converged on steatite, which offered a crucial
advantage: when fired at high temperature, steatite becomes hard and glassy, allowing
seals to be used for longer term, perhaps persisting across repeated trade cycles.
This firing innovation provided the first long-term contract with perhaps an
anti-counterfeiting mechanism. Steatite seals are indicative of centralization
and standardization in trade practices, reflecting the growing complexity of economic
interactions.

Additionally, seals and sealings were small and portable---ideal for record
keeping in pre-literate communities and easily carried. At this stage, perhaps
the number of sealings at the receiving end could serve as a rudimentary record
of incoming goods between counterparties.  Parties could meet seasonally to
reconcile records by simply matching the number of sealings each party held
and settle accounts.

\subsubsection{Animal and Plant Motifs (Mature Harappan, c.\ 2600--2200 BCE): Expansion Beyond Family Networks}

As trade scaled beyond local, kinship-based networks, the geometric system reached a
critical bottleneck: geometric patterns could not reliably distinguish between different
traders or trading houses. Shippers, transport agents, and their workers would
have had difficulty determining where packages should be routed. The system
required clearer identity markers.

The solution was the adoption of real totemic animal and plant motifs---bull,
elephant, rhinoceros, tiger, gharial, and various plant designs. These motifs
carried cultural resonance within their respective ethnic or regional communities,
functioning as recognizable emblems for specific trading houses and families. Each could
be assigned to a guild or merchant collective, serving as a brand identifier and
indexing product lines or specialization.

However, this system too faced a critical scaling problem: the number of real
animals, plants, and their recognizable variants is finite. More importantly,
as the trade network expanded to draw merchants from diverse cultural backgrounds,
the use of totemic animals could create political issues. Additionally,
coordinated seal production was needed to prevent issues where different parties
could choose the same iconography, and to establish something neutral to avoid trade
frictions related to routing and handling.

\subsubsection{The Invented Unicorn (c.\ 2600--1900 BCE): Breaking the Cardinality Ceiling}

The unicorn---a fantastic creature with bull-like body, single forward horn, slender
proportions, and no natural prototype---appears suddenly in the Mature Era and dominates
approximately 60\% of seals thereafter. This motif was likely an administrative or
commercial invention, born from the convergence of two pressures: the exhaustion of real
animal variants and the need for a politically neutral symbol. Unlike real fauna and flora
that carried totemic significance to particular cultural groups, the unicorn belonged to
none. It functioned as a compromise animal---a creature that could represent network
membership without according primacy to any single cultural or totemic group.

The invented unicorn and the stand broke the cardinality ceiling: the repertoire
became extensible beyond the finite inventory of recognizable real animals and
plants. More importantly, the unicorn signified membership in the confederation itself,
analogous to a modern corporate logo or franchise mark. Its universality conveyed that 
a seal-user belonged to the confederation united across the Indus valley—not because of 
ethnic, regional, or totemic affiliation, but through voluntary commercial participation 
and confederal membership. Each merchant guild retained its distinct offering stand as a 
marker of individual identity and guild autonomy. The exceptional diversity of offering 
stands in the corpus---139 distinct types across 19 sites---provides direct evidence of 
the confederation's scale and inclusive membership structure: each stand variant 
represents a merchant guild or trading collective admitted to the confederation, making 
the stand repertoire itself a registry of the confederation's expansive, pluralistic 
architecture where autonomy and unified standardization coexist.

\subsection{Archaeological Evidence for the Trade}

The seal evolution documented by \citet{kenoyer2010inscribed} and colleagues is grounded
in stratigraphic and contextual evidence. But to understand how this system actually
operated—where seals were deployed, how they were managed, and what operational patterns
they reveal—we turn to detailed archaeological documentation of seal distribution and use
contexts. Based on this evidence, we can propose a plausible operational architecture
of the Indus commercial network.

\subsubsection{No Anthroponyms and Toponyms}

\citet{mukhopadhyay2023} establishes that Indus script inscriptions encode neither
anthroponyms nor toponyms. Our statistical findings confirm this conclusion.
This is crucial: it means the seal script system could not identify individuals
or places by name. Instead, identity had to be encoded through non-linguistic means:
the unicorn motif (commercial network marker), the offering stand variant (guild
affiliation), and the accompanying attribute bundle (confirmation checksum).

\subsubsection{Seal Distribution and Association}

Archaeological evidence confirms that seals served trade related logistic functions.
\citet{mukhopadhyay2023} documents that inscribed seals concentrate in functionally
specific contexts: near city gates, craft workshops, and public buildings, often
alongside standardized Indus weights. Sealings appear attached to storage containers
and warehouse locking systems. These contexts are uniformly commercial and
logistic—gateways for traffic control, workshops for production control, warehouses
for access control. The seals functioned to enforce ``rules involving taxation,
trade/craft control, commodity control and access control'' \citep{mukhopadhyay2023}.
They were not decorative, not ritual, not administrative records (which would require
centralized archiving). They were tokens deployed at points of commercial
decision-making.

\subsubsection{Trade Control Architecture}

\citet{frenez2005lothal} provides the most detailed archaeological documentation of
sealing practices from Lothal. His work documents 93 individual sealings recovered
from a warehouse context, with 70 concentrated in a single square meter (Block C,
SRG3, Layers 2–3), representing a logistical checkpoint. He notes a pattern:
across the Indus civilization, the seal-to-sealing ratio is approximately 10:1
(over ~3,700 seals known, but only ~200 tablets and ~130 sealings total). Frenez notes
that this ratio contrasts sharply with Mesopotamian evidence, where sealings are
nearly as abundant as seals and are maintained in administrative archives. The Lothal
cache itself was preserved accidentally by fire---the only reason such a sealing archive
survives in the archaeological record. This scarcity of sealings raises a fundamental
operational question: why would seals be preserved and reused, while sealings appear
only rarely and must be reconstructed from accidental preservation?

\citet{fuls2023catalog} cataloged Mound F at Harappa and documents a large
inventory: 173 square seals, 36 rectangular seals, 164 bas-relief tablets, and
215 inscribed tablets. Kenoyer and Meadow interpret this as warehouse storage.
Crucially, the Mound F cache contains no sealings—only the permanent tokens
(seals and tablets). This pattern aligns with Frenez's findings: seals and
tablets persist in the archaeological record, while sealings are ephemeral.
Together, Frenez and Fuls provide evidence consistent with a trade control
architecture.

\section{Core Inferences from Statistics and Archaeology}
\label{sec:core_inferences}

The statistical patterns of the Unicorn Seal System (USS) are consistent with a
self-organizing network rather than a centrally controlled administrative
system. We take this as a point of departure for understanding the Harappan
civilization as a trade-driven commercial network spanning a vast region. The
statistics alone do not establish this reading. It rests on the trade contexts in
which seals and sealings are found, and on the absence of rulers or religious
imagery in the archaeological record, which makes a commercial interpretation the
most parsimonious one.

In the scale-free network models of \citet{barabasi1999emergence} and
\citet{bianconi2001competition} discussed earlier, new entrants preferentially
link to established hubs rather than to peers at random. This produces a
self-reinforcing dynamic: successful merchants attract more partners, which
increases their reach and reputation (fitness) and attracts further
partnerships. Merchants seeking to minimize risk and maximize access to reliable
partners would naturally converge on established hubs. Aggregated over hundreds
of merchants and many generations, such decentralized choices can produce a
scale-free structure, with a small number of hub guilds linking many peripheral
traders. Such a topology requires no central authority; it can emerge from
individual commercial choices and network effects
\citep{katz1985network,arthur1989competing}.

Further clarification, in network science terms, when seals are interpreted as nodes
representing commercial entities or trading houses, a direct link or shared interaction
implies an active trade relationship within a Barab{\'a}si-Albert (BA) or
Bianconi--Barab{\'a}si (BB) network topology.

\subsection{Claim \#1: Offering Stand as Commercial Entity Emblem}

We argue that each unique offering stand type (139) is the emblem of a
persistent group engaged in trade, an identifier that maintained relationships
across sites and time within the USS. The key fact that it is visually seperate
on the seal face, is the most diverse non-animal motif and it follows power-law
distribution that conforms to a node in a trade network. Over the roughly
700 years of the Mature Harappan period, the same stand types recur across sites,
which points to persistent group identities rather than one-off artisanal choices. We
read these groups as commercial entities because stands occur in trade contexts and
because no competing religious or royal interpretation is visible in the record. Other
group-level interpretations, such as kin or clan marks, could produce similar
distributions, so this reading rests on the archaeological context rather than on
the statistics.

\subsection{Claim \#2: Unicorn Seal System as a Winning Standard or Technology}

The predominance of the USS, together with its power-law form, suggests a seal-based
standared or technology that came to prevail in the Mature phase of the Harappan
civilization, in accordance with network-externality theory. The diachronic data are
consistent with increasing participation by commercial entities within the USS (4, 45,
and 93 seals across the three periods).  \citet{arthur1989competing} describe the
process: ``when two or more increasing-returns technologies `compete' for a `market'
of potential adopters, insignificant events may by chance give one of them an initial
advantage in adoptions. This technology may then improve more than the others, so it
may appeal to a wider proportion of potential adopters. It may therefore become
further adopted and further improved.'' The USS is such a standard that was adopted
by more and more of the commercial entities described in the previous claim.

\subsection{Claim \#3: Commercial Entity Participation in Trade Contracts or Agreements}

As described above, commercial entities participated in trade links connecting distant
sites across the Harappan trade network. Each of these commercial entity participated in
forming a link with another entity with top few of them becoming the hubs of the network
in accordance with network theory.  We propose that a pair of such seals, one for each
party, would serve as the record of an agreement or contract between them in the network.
We infer this from network science as well as trade related context where seals are found.
We cannot identify specific trading partners archaeologically, but the network science
is consistent with the such a preferential linkage. The wide distribution of some
top stand types across sites indicates the breadth of that entity's trading
participation. Some stands appear at many dispersed sites, indicating extensive
trading links; others appear at a single location, indicating limited reach or
peripheral status.

\subsection{Claim \#4: Coordinating Institutional Requirements}

The cross-site agreement between stands and attributes (Table~\ref{tab:cross_site})
indicates a shared convention for pairing stand with animal features, maintained across
hub cities and distant settlements. A convention of this kind suggests some form of
coordination. The agreement is partial
rather than uniform, so the evidence supports a shared standard with variation, not
strict enforcement. Hub guilds may have developed the USS cooperatively as a trade
and logistics innovation \citep{katz1985network}, since they had the resources to
manufacture seals, administer the system, and enable smaller traders to
participate. Singleton stands, found at only one site, plausibly lacked such
resources, yet they carry the same convention; how they obtained properly made
and validated seals is the coordination question. A trained, mobile workshop
tradition could also produce this pattern, but given the trade context a
coordinating institution is the more parsimonious explanation. This is the
another hypothesis of the paper, and it is an inference from the statistics
together with the archaeology, not a result of the statistics alone.

Such an institution or federation would need to:

\begin{itemize}
    \item validate new entrants seeking seals, prevent counterfeiting, and keep
    each stand unique so that it identifies its holder durably;
    \item provide seal infrastructure to independent producers, through
    manufacturing, logistics coordination, or licensing;
    \item ensure protocol compliance by enforcing standards across participants;
    \item resolve disputes arising from contractual arrangements.
\end{itemize}

The evidence does not specify whether these functions were concentrated in a
single organization, distributed among guilds, or exercised through a commercial
federation. The most frequent stands (\#16, \#1, \#30, \#32, \#44) may have had
disproportionate influence in setting standards, but the long tail of smaller
participants indicates that the system was not a closed group. We hypothesize that
the Unicorn Logistics Confederation (ULC), described in the model developed
below, is such an institution, and that its role helps explain the Unicorn's
dominance in seal iconography. The following section examines whether the
archaeological record preserves the operational traces such a system should leave.

\subsection{Claim \#5: Scale-Free Signature and self organization}

A heavy-tailed, power-law-like signature appears at several levels of Harappan
glyptic art: in the overall seal motif (Unicorn and others) distribution, within
the Unicorn Seal System itself with stand types, and in site-level counts. This is
suggestive of self-organization as an organizing principle in the Harappan
civilization that recurs at many levels of sealing technology.

\section{Seals, Tablets and Sealings: Asynchronous Barter System and Clearing Exchange}

The Harappan trade network spanned more than a million square kilometers, the key
challenge is one of logistics of a variety of packages type to be safely conveyed
multi-modally as well as via multiple parties spanning diverse regions of the Indus
area. We propose that the Harappan sealing system played a central role in
coordinating this complex logistics as well as accounting within the trade network.

The statistical patterns and archaeological contexts above motivate a
hypothetical model organized around three coordination challenges:
\begin{itemize}
    \item \textbf{Routing across distance and language:} Long-distance exchange
    could involve multiple transport modes and linguistic communities. We propose
    that the USS's two-layer information system combined rapid visual
    identification and routing with more detailed, specialist-readable information.
    \item \textbf{Reconciliation over time:} Because shipments were asynchronous,
    merchants needed to review which consignments had arrived, which remained
    outstanding, and how obligations balanced. Recurring meetings on an agreed
    schedule could provide a time and place for reconciliation, renewed agreements,
    and dispute resolution.
    \item \textbf{Secure handover between parties:} A proposed protocol had to
    preserve identity and accountability as goods passed from Party A through
    coordinating logistics providers to Party B, and in reciprocal exchange from
    B to A. We model the ULC as coordinating that process, without assuming the
    archaeological record proves such an institution existed.
\end{itemize}
These challenges organize the hypotheses developed below: the USS's two-layer
information architecture, periodic meetings, the Asynchronous Peer-to-Peer Barter
System (AP2PBS), and the Unicorn Logistics Confederation (ULC). The archaeological
record does not directly document these institutions or practices.

\subsection{Two-Layer Information Architecture}

We propose that the statistical correlation within the USS is consistent with
it being designed in two distinct layers:

\begin{table}[H]
\centering
\caption{The seal's two-layer information architecture.}
\label{tab:two_layer_inferences}
\begin{tabular}{@{}lll@{}}
\toprule
 & \textbf{Layer 1 --- Iconography (Stand)} & 
   \textbf{Layer 2 --- Script} \\
\midrule
Audience & Package sorters, dockworkers, & Guild members, scribes, \\
 & gatekeepers, handlers of goods & trade specialists \\
Content & Guild emblem (stand) + visual & Product detail, quantities, \\
 & confirmation checksum & types, measures \\
 & (neck/halter etc.) & \\
Function & Rapid visual verification of & Precise record-keeping and \\
 & guild identity and authenticity & contractual terms \\
Skill required & None—readable without & High—specialist-readable, \\
 & literacy across regions & needed only at origin/destination \\
\bottomrule
\end{tabular}
\end{table}

This two-layer design solves a fundamental Bronze Age logistics problem:
authenticating goods across vast geography without literacy while enabling
reliable package routing and claiming without visible centralized administration.
Layer 1 is universal and fast and hard to forge; Layer 2 is secure but requires
specialists. This dual-layer system specifically was iteratively arrived at
to solve the multi-modal logistic challenge of long distance barter.

\subsubsection{Attribute Correlations: A Visual Confirmation Protocol}
\label{subsec:correlations}

Packages were secured with clay impressed by the seal. Intermediaries at
way-stations, caravan stops and ports had to validate package identity from
sealings for onward routing, across linguistic and cultural zones. When a sealing
was damaged or illegible, handlers could fall back on visual pattern-matching
rather than script; a package would also carry multiple impressions for
redundancy. We propose that the offering stand served as the \textit{primary
visual identifier}, quickly distinguishing one merchant's packages from another's,
and that the characteristic attributes (neck, halter, ear, etc.) paired with it
formed a \textit{confirmation protocol}. A handler would check the attributes
against the expected bundle for that stand---``Does this Stand 16 display the
expected Halter variants? Yes, so route to Vessel A''---allowing rapid sorting
where intermediaries could not read script. A constrained bundle also deters
fraud, since an imposter would struggle to reproduce all of it. At origin and
destination, parties presumably knew the script and could verify seals through
several channels; the visual bundle was redundant there but reinforced identity.
The stand-attribute correlations and the cross-site agreement reported earlier are
consistent with such a convention, but do not show that it was used this way.

\subsection{Periodic Meetings: Reconciliation, Renewal, and Innovation}

In this trade driven model, periodic meetings brought merchants together to manage
the parts of trade that could not be completed through asynchronous shipment alone.
At recurring gatherings, participants could reconcile outstanding obligations, and
resolve disagreements over delayed, damaged, or disputed consignments. By netting
reciprocal claims, they could close one cycle of exchange.

These gatherings would also provide occasions to negotiate new agreements and revise
existing terms. Merchants could respond to changing supplies and demand by setting
new prices or exchange ratios, agreeing on quantities and quality standards, and
introducing newly available products to trading partners. Repeated face-to-face
contact could renew relationships and trust, while shared decisions could maintain
compatible practices across the network. In this sense, meetings were not only
settlement events but also forums for commercial coordination and innovation.

The archaeological record does not identify this protocol or institution,
so their periodicity and functions remain inferential.

\subsection{The Unicorn Logistics Confederation (ULC): Institutional Functions}

The ULC's value proposition rested on several crucial functions that individual
merchants could not provide independently without mutual coordination:

\begin{itemize}
\item \textbf{Guaranteed weight standardization:} The ULC maintained and distributed uniform
weights (hexahedral chert stones) across all major sites. This ensured that when Party A
contracted to send ``100 standard units of carnelian,'' Party B could verify the quantity
using identical weights at their location. Without standardized weights, transactions
would fail due to disputes over quantity.

\item \textbf{Controlled seal manufacture and authentication:} The ULC controlled seal
production and authorized them. A logistic entity encountering an unfamiliar
seal could trust that if the ULC had authorized it, it belonged to a recognized guild.
The production of seals may have been decentralized but still following ULC standards
for stand and attribute bundle as well as repertoire of script.

\item \textbf{Dispute settlement and arbitration:} When discrepancies arose—missing
goods, damaged shipments, unmet contractual terms—the ULC provided neutral arbitration
at seasonal settlement conferences. Parties brought their evidence (tablets with
discrepancy markings, sealings), and ULC arbiters negotiated resolutions binding on all
parties.

\item \textbf{Secure delivery guarantee:} The ULC physically transported goods, provided
waypoint security, and vouched for chain-of-custody. If goods were lost or stolen in
transit, the ULC bore liability (subject to proof of delivery via sealings).

\item \textbf{Counterparty risk underwriting:} This was the crucial function. Before the
ULC, a merchant sending goods faced absolute risk: what if the counterparty never sent
their promised goods back? The ULC solved this by vouching for both parties' good faith.
By placing its reputation behind every transaction, the ULC reduced counterparty risk to
manageable levels. Without this underwriting, stranger-based trade over such vast
distances would have been economically irrational.

\item \textbf{Multi-modal logistics coordination:} The ULC maintained expertise across
diverse transport modes---carts, pack animals, river boats, maritime vessels---and knew
optimal routing through different seasons and ecological zones. Individual merchants
need not master all this knowledge; it would be impractical to do so individually.
The ULC provided this expertise.
\end{itemize}

These functions explain why any guild would participate in the ULC---these services
are inferential and not directly observable. The form of compensation remains
uncertain: participation may have involved fees, licensing payments, membership
contributions, reciprocal obligations, or other benefits exchanged for access to
the ULC's services. Whatever the precise arrangement, the ULC solved coordination
problems that isolated merchants could not solve alone, making participation valuable
even for guilds that could not maintain independent long-distance infrastructure.
Archaeological evidence from Kotada Bhadli (Kachchh) identifies possible
caravanserai infrastructure, suggesting that such way-stations and logistics nodes
existed to support long-distance trade networks and lending support to the hypothesis
that a coordinating institution like the ULC would have maintained physical sites
for goods staging, waypoint verification, and merchant assembly
\citep{shirvalkar2025caravanserai}.

\subsection{Operational Model: Asynchronous Peer-2-Peer Barter System (AP2PBS)}

We propose an operational model that could explain this evidence: the
Asynchronous Peer-to-Peer Barter Settlement (AP2PBS) protocol. AP2PBS is a
merchant-developed protocol that emerged from the collaborative efforts of multiple
guilds seeking to coordinate trade across unprecedented distances and cultural
boundaries. The proposed AP2PBS protocol explains how seals, tablets,
and sealings were used to actually handle the operational lifecycle of long-distance
barter and settlement.  AP2PBS provided more than a sequence for moving goods: it
anchored trust across an almost million-square-kilometre geography.  AP2PBS made
distant exchange operationally legible; the ULC made participation in that system
credible.

The following five phases outline how merchants, aided by the ULC, conducted
asynchronous peer-to-peer barter across vast distances.

\subsubsection{Phase 1: Conference—Initial Agreement}

Merchants A and B meet face-to-face (at a conference venue, market hub, or port) to
negotiate a future barter transaction:

\begin{itemize}
\item \textbf{Negotiation of terms:} Merchants agree on goods, quantities, and
quality standards. Example: ``A will send 100 carnelian beads of grade X to B;
B will send 50 grain sacks of quality Y to A.''

\item \textbf{Agreement inscription on seals:} The agreed terms are carved into
unique seals. Each merchant designs or commissions a seal bearing their guild's
unicorn + offering stand + confirmation attributes, with script encoding the
product and quantity metadata.

\item \textbf{Tablet exchange:} Each merchant produces two clay tablets impressed
with their own seal (which have their guild's identifying attributes) and gives both
tablets to the other party. These tablets are ``claim vouchers'' as well as ``routing
manifest''. The ``claim voucher'' B is for A to claim goods from B delivered by the ULC.
The ``routing manifest'' B is for A to send their goods with sealing A which the ULC will
route to B.  This is the simplest form of a peer-to-peer barter arrangement.

They could be multiple such pairs of tablets each indicating mutual obligations that
each party could keep as a record of their obligations across multiple cycles.
We note here that these tablets may have perforations, allowing them to be strung
and attached to the packages for usage as a routing manifest.

\item \textbf{Parties depart:} Merchants return to their home settlements. The
tablets are their portable proof of obligation. Without these tablets, the ULC will
not release goods at the destination.
\end{itemize}

\subsubsection{Phase 2: Shipment—Origin Verification and Packing}

Party A produces the agreed goods and prepares them for transit:

\begin{itemize}
\item \textbf{ULC witnessing and verification:} ULC agents examine the goods, verify
their quality and quantity against standardized weights, and confirm the match with
the agreed terms carved on A's seal.

\item \textbf{Package sealing:} ULC agents wrap the package with cordage and seal
it with Sealing A—an unfired clay lump impressed with A's seal. The sealing proves
that the goods match the agreement and provides tamper-proof evidence of package
integrity.

\item \textbf{Tablet handover:} Party A hands Tablet B (bearing Party B's seal
impression) to the ULC agents. This tablet becomes the ULC's routing document—it
specifies the final destination and recipient. It likely was strung to the package
for routing purposes. It would have to be strung to the package throughout its journey
separate from the sealing itself to maintain the integrity of both the routing
information and the package seal.

\item \textbf{Departure:} The package (bearing Sealing A) and Tablet B (held by
ULC agents) embark on transit. They will travel together to Party B's location
but remain physically separate.

\end{itemize}

\subsubsection{Phase 3: Network Routing and Intermodal Transit—ULC Management}

In Phase 3, the package (bearing Sealing A) and Tablet B (the routing manifest)
traverse the network from origin to destination under ULC management. This phase
is the operational heart of AP2PBS—where the stand + confirmation bundle,
statistically demonstrated throughout our analysis as the core of the seal
system, becomes a functional logistics tool.

\begin{itemize}

\item \textbf{Stand + Confirmation bundle as transactional identity:} The seal on
Tablet B --the unicorn plus offering stand plus attributes (Cram{\'e}r's
V \textgreater 0.45 correlation)--constitutes the confirmation bundle. This bundle
is the primary identifier across the entire network. Each guild maintains a distinctive
stand composition and attribute set, visually recognizable at any waypoint across all
sites and languages. This confirmation bundle is not decorative; it is logistics
infrastructure coordinated by the ULC. It allows ULC handlers at multilingual waypoints
to authenticate the shipment and forward it without reading script.

\item \textbf{Permanent ULC site agents as network nodes:} At each major site
(Harappa, Mohenjo-daro, Dholavira, Lothal, Shortugai, etc.), the ULC maintains
a permanently stationed site agent. This agent is not a visiting merchant but
a professional logistics operator—multilingual, culturally versed in local
practices, and embedded in the local community. The site agent's
responsibilities form the operational backbone of the system: receiving
shipments, verifying seals and tablets, holding goods in secure storage
(warehouse), releasing consignments to outbound carriers or to the final
recipient, managing weight standards, witnessing discrepancies, and gathering
market intelligence on local trade flows and merchant behavior.

\item \textbf{Market intelligence and counterparty monitoring:} The permanent
site agent is the network's eyes and ears. By observing which merchants are
sending goods to which destinations, in what quantities and commodities, the
agent gathers intelligence on market patterns, seasonal flows, and merchant
relationships. This information flows back to senior ULC coordinators at hub
sites. If a merchant suddenly fails to send expected goods, if patterns shift
dramatically, or if disputes arise, the site agent's observations provide
evidence for ULC arbitration (Phase 5). This function also serves an
early-warning purpose: if a guild begins to show signs of financial distress
or unreliability, the site agent can flag it, allowing the ULC to reduce
counterparty risk exposure.
\end{itemize}

\textbf{Link to ULC Functions:} Phase 3 operationalizes all ULC functions
from Section 7.2. The permanent site agents enforce weight standardization
( \textbf{Function 1}), authenticate seals via the confirmation bundle
( \textbf{Function 2}), document discrepancies as future arbitration
evidence ( \textbf{Function 3}), manage multimodal transport routing
( \textbf{Function 6}), and gather the intelligence needed to underwrite
counterparty risk ( \textbf{Function 5}).

\subsubsection{Phase 4: Two-Key Delivery Handover—Mutual Authentication}

Party B (or their agent) appears to claim the goods. Dual authentication
ensures both parties are legitimate:

\begin{itemize}
\item \textbf{Party B's tablet check:} Party B presents Tablet A as well
as their own seal to ULC agents. ULC agent matches the presented Tablet A
with Sealing A and confirms Party B's seal against the routing authority token
(Tablet B) that Party A handed to the ULC.

\item \textbf{Package sealing comparison:} ULC agents compare the seal
impression on Tablet A (presented by Party B) with the impression on
Sealing A (the package seal). These must match. If they match, Party B
is legitimate and authorized to receive the goods.

\item \textbf{Mutual authentication:} Simultaneously, ULC agents present
Tablet B to Party B. Party B examines this tablet and confirms it bears
their own seal—proof that the ULC agent is the genuine, authorized carrier
sent by the original agreement.

\item \textbf{Re-weighing and discrepancy marking:} ULC agents open the
package in front of Party B as an official witness. They re-weigh the
contents against standard weights. If shortages or damage are found,
Party B marks these discrepancies directly onto the back of Tablet A
using a stylus (incising scratches to indicate units missing or damaged).

\item \textbf{Sealing A as receipt:} ULC agents take Sealing A (only
one, though packages may be stamped with multiple sealings bearing A's seal for
redundancy), as proof that delivery occurred as contracted. Party B
receives the goods and keeps Tablet A (with or without discrepancy markings).
Tablet A is now a completed record of one leg of the transaction by Party A.
\end{itemize}

Similarly, Party B would complete asynchronously their leg of the transaction.

\subsubsection{Phase 5: Reconciliation and Settlement—Periodic Clearing}

At the turning of seasons, all principals reconvene at a central clearing venue:

\begin{itemize}
\item \textbf{Token assembly:} Merchants bring their physical evidence.
Party A brings Tablet B (with or without discrepancy markings); Party B brings
Tablet A (with or without discrepancy markings). ULC agents bring delivery
sealings.

\item \textbf{Obligation reconciliation:} Parties lay out tokens and reconcile.
If Party A's Tablet B shows no discrepancies and Party B's Tablet A matches the
goods received, the obligations cancel. If discrepancies exist (missing goods,
damage), parties negotiate adjustments: partial refunds, compensatory goods
in next cycle, or arbitration awards.

\item \textbf{Deliberate destruction:} Once multi-party debts are
squared and balanced to zero, the physical tokens (tablets and sealings) lose
all legal value. To prevent fraud and double-counting, they are intentionally
destroyed—broken, discarded, or thrown into architectural fills and refuse
deposits at the settlement venue. At this time if agreements breakdown or
there is need for newer agreements, they destroy old seals.

\item \textbf{Network reset:} The transactional cycle closes. Accounts are
cleared. The platform resets for the next seasonal wave of trade. Parties
depart, ready to negotiate new agreements.  Phase 1 of the next cycle then
begins with the creation of new agreements and corresponding seals and
tablets.
\end{itemize}

\subsection{Archaeological Evidence: Reinterpreted according to AP2PBS}

\subsubsection{Seal Distribution and Guild Identities}

\citet{mukhopadhyay2023} documents seal contexts across the Indus civilization
systematically. Her observations, compared with the AP2PBS model, show convergence
on core points and demonstrate where AP2PBS adds operational detail. She proposes
administrative and taxation functions; we determine that commercial function is
a more parsimonious explanation, given the lack of evidence for an extractive or
centralized state in the archaeological record. The following table presents this
comparison:

\begin{table}[htbp]
\centering
\caption{Mukhopadhyay's archaeological observations vs. AP2PBS operational interpretation.
Mukhopadhyay establishes WHAT seals, tablets, and sealings were found and WHERE. AP2PBS
explains HOW they functioned in practice.}
\label{tab:mukhopadhyay_abs_comparison}
\scriptsize
\setlength{\tabcolsep}{2.5pt}
\renewcommand{\arraystretch}{0.8}
\begin{tabular}{@{}p{3.4cm}p{4.8cm}p{6.8cm}@{}}
\toprule
\textbf{Artifact/Context} & \textbf{Mukhopadhyay's Finding} & \textbf{AP2PBS Operational Role} \\
\midrule
Seals (distribution) & Concentrated at gates, workshops, storage facilities
with standardized weights & Permanent guild identity tools; ULC agents carry seals
at each waypoint for authentication \\
\midrule
Identical iconography or inscriptions on seals, tablets and sealings at distant
sites & Standardized regulation, permits, licensing & Asynchronous logistics: Seals
at origin; tablets at destination (matching pairs); AP2PBS phases; \\
\midrule
Identical iconography with differing inscriptions on seals & Officials/artisans/
merchants of same guilds/rulers/governing-bodies controlled/practiced differing
crafts/trade & We agree but frame it as commercial differentiation of products \\
\midrule
Differing iconography with similar inscriptions on seals & Officials/artisans/
merchants of different guilds/rulers/governing-bodies controlled/practiced
similar crafts/trade & We agree but frame it as commercial competition for
same products \\
\midrule
Inscribed sealing-pendants of Kanmer & Conjectured to be passports and
gate-passes & Asynchronous logistics: Seals at origin; tablets and sealings at
destination (matching pairs); AP2PBS phases; additionally, these may have served
as high-security escorts for valuable consignments; or perforated tablets
could have been used as the routing tablet \\
\midrule
Similar tablets at distant sites & Bear seal impressions on obverse; smooth
backs or with numerical-metrological notations. License and tax marking &
Routing documents (Tablet B) held by ULC; claim vouchers (Tablet A/B) for
counterparties; smooth backs allow discrepancy marking; see AP2PBS phases \\
\midrule
Seal + weight association & Seals and weights co-occur at gates/workshops
& ULC agents verify goods weight against standard measures before sealing
(Phase 2) and after delivery (Phase 4) \\
\midrule
Script (no names/places) & Script encodes neither anthroponyms nor toponyms
& Script encodes product metadata only; guild identity via seal, not written
name; routing via tablet, not written address \\
\midrule
Iconography & Mukhopadhyay documents seal contexts systematically, proposing administrative and taxation functions & Permanent
guild identity. We confirm this via our statistical data - with commercial
entity interpretation \\
\midrule
Disappearance after IVC decline & Seals/tablets/sealings cease ca. \ 1900 BCE
& Trade system collapse: without ULC coordination, AP2PBS has no operational
basis \\
\bottomrule
\end{tabular}
\end{table}

\FloatBarrier

\subsubsection{Frenez's Warehouse Evidence: Logistics and Operational Patterns}

\citet{frenez2005lothal} documents 93 sealings from Lothal's warehouse excavation
context in detail.  His observations of sealing types, container associations,
and spatial clustering reveal patterns consistent with asynchronous logistics.
We interpret Frenez's empirical data through the AP2PBS operational model to
show how these warehouse patterns could have functioned in long-distance
trade coordination. The following table maps Frenez's documented findings to
AP2PBS phases:

\begin{table}[H]
\centering
\caption{Frenez's Lothal sealing data interpreted through AP2PBS operational model.
Frenez documents archaeological patterns; AP2PBS explains their functional meaning
for trade logistics.}
\label{tab:frenez_abs_comparison}
\begin{tabular}{@{}p{4cm}p{6cm}p{6cm}@{}}
\toprule
\textbf{Sealing Pattern} & \textbf{Frenez's Observation} & \textbf{AP2PBS
Interpretation (Phases 1–5)} \\
\midrule
Multi-seal sealings (up to 5 seals per sealing) & L-189, L-190 bear 4 seals
each & Multiple guilds' packages staged together in warehouse pending
consolidated shipment (Phase 3 waypoint staging) \\
\midrule
Dense warehouse cluster (70 in one square meter) & Sealings concentrated in
Block C, SRG3, Layers 2–3 & Logistical queue: outbound shipments awaiting
transport; temporary accumulation before dispatch (Phase 3) \\
\midrule
Diverse containers (wooden boxes, pottery, sacks, canes) & L-161–172 (boxes),
L-177–181 (pottery), L-126 (sacks), L-134–140 (canes) & Different commodity
types require different containers; ULC managed multimodal packing
(Phase 2–3) \\
\midrule
Fingernail tallies on boxes & L-161–172 bear 1–6 tally marks & Discrepancy
records: units missing, damaged, or reconciled at delivery (Phase 4 re-weighing) \\
\midrule
Seal reuse (13 seals, 76 impressions) & High-frequency seals (L-6: 16
impressions; L-37: 12 impressions) & Durable guild seals reused across multiple
shipments; contrasts with one-time-use sealings (Phase 2 creation,
Phase 4 handover) \\
\midrule
Pegs-on-wall sealings & L-179, L-206 with vegetal fiber ropes & Storage room
locks for batched goods: temporary security during Phase 3 warehouse staging \\
\midrule
Unfired sealings + accidental fire preservation & 70 sealings preserved only
by fire at Lothal & Sealings are disposable: routinely destroyed after use
(Phase 5 clearing); fire preserved only rare intact examples \\
\midrule
10:1 seal-to-sealing ratio (across Indus) & ~5,000 seals known vs. \ ~130–140
sealings & Seals are permanent tools (reused across many cycles); sealings are
temporary labels (created per-shipment, destroyed at settlement) \\
\bottomrule
\end{tabular}
\end{table}

The Lothal sealings were also classified with respect to their find spots
to be from the "warehouse" context. It is also noted that this area was near the
place where boats docked - again reinforcing the trade context of these seal,
sealing, and tablet finds. \citet{frenez2024} classifies most of the finds
to the "warehouse," where he identifies multi-stamped sealings - we interpret these
as packages awaiting shipment before loading onto boats that would eventually
carry them.

\subsection{Predictions and Confirmations of AP2PBS}

The AP2PBS model generates a series of testable predictions.

\subsubsection{Stand and Script commonality}

The seals bearing the same stand iconography could be tested to see if they have
common script.  It would suggest that the commercial entity specializes in a
particular class of goods or trade activity, perhaps with multiple partners.

If a particular stand type had a large inventory of scripts, it would likely indicate
the breadth of products or transaction that guild had.

\subsubsection{Trade Linkages}

Identical inscriptions or iconography on seals and tablets found at sites separated
by hundreds of kilometers are consistent with two parties holding complementary
routing and claim tablets at opposite ends of a trade arrangement. In the AP2PBS
model, one party's tablet authorizes a claim or provides routing authority for the
other party, while the corresponding seal or tablet authenticates the consignment
at the other end. Such a match links the two sites by at least one leg of the trade,
without by itself demonstrating a complete route, a single commercial owner, or
the full sequence of intermediaries between them.

\subsubsection{Venues for Periodic Settlement and Conference}

AP2PBS barter arrangements require periodic settlement meetings in a large, non-residential
structures capable of hosting multiple merchants for periodic conference and settlement
would be expected. Functional venues could include spacious halls with aligned seating,
and nearby storage areas for goods staging. Such venues would not be temples or palaces
but explicitly commercial infrastructure.

Pillared hall, as found in Mohenjo-Daro, may have served this function, providing a
covered space for merchants to reconcile tablets and sealings in addition to negotiate,
and settle disputes.

\subsubsection{Scarcity of Tablets and Sealings Relative to Seals}

Because settled pairs are deliberately destroyed, tablets and sealings should be far
less common than seals (which are permanent long term contracts). \citet{frenez2024}
reports the ratio of 10:1 seals to sealings. Over ~3,700 seals are known, but only
approximately 200 tablets and a similar number of sealings—Lothal being the largest
single deposit. This ratio is exactly what AP2PBS predicts: reusable seals create
thousands of tokens; disposable tablets and sealings create only dozens of
archaeological traces.

\subsection{Why AP2PBS Explains the Unicorn Dominance}

As velocity of trade increased, the network expanded. More merchants, more guilds, more
products, more trade links.  The unicorn's dominance is therefore a direct consequence
of AP2PBS and ULC driving network expansion.

This functional trade related innovation is consistent with what the quantitative evidence
documents.  The diachronic data reveal a distinctive statistical signature: rapid adoption
cascade followed by plateau, consistent with preferential-attachment dynamics in
an expanding network.  The network's observed growth trajectory is the expected outcome
hen a decentralized coalition of independent traders adopts a politically neutral,
scalable trade protocol.

\section{A Functional Reinterpretation of Harappan Civic Architecture}

\subsection{Form Follows Function: Trade Infrastructure, Not Temples and Palaces}

The monumental buildings of Mohenjo-daro have often been interpreted through
priest-king paradigms derived from other Bronze Age civilizations. Earlier
excavators and subsequent summaries variously described the Pillared Hall as an
``assembly hall'' or ``prayer hall,'' the Great Bath as a temple or ritual
bathing space, and the so-called ``Great Granary'' as a facility for
administrative or grain storage. These labels, however, remain functional
hypotheses rather than demonstrated identifications. More recent archaeological
interpretations have questioned the older palace, temple, and granary models
and have emphasized the uncertainty of the buildings' original functions.

The absence of securely identified palaces, royal tombs, monumental royal
inscriptions, and clear priestly iconography makes a centralized priest-king
interpretation difficult to sustain. At the same time, the Harappan record is
rich in evidence for organized production and exchange: seals, standardized
weights, tablets, sealings, warehouses, workshops, gateways, and extensive
transport infrastructure. Read together, these materials support interpreting
Mohenjo-daro's monumental structures within the wider commercial and logistical
system rather than assigning them inherited political or ritual functions.

Applying Louis Sullivan's dictum---\textbf{form ever follows function}---we
therefore interpret these structures as possible components of the operational
infrastructure of AP2PBS, or of a comparable long-distance barter system. This
interpretation remains a reconstruction rather than a direct archaeological
identification, but it explains the buildings within the broader evidence for
distributed commercial coordination.

\subsection{Three Functional Venues}

\subsubsection{The Pillared Hall: Phase 5 Settlement Chamber}

\textbf{Physical evidence:}
\begin{itemize}
\item 27.5 metre square; 20 brick pillars in 4×5 rows, spaced 9'7'' ± 3 inches (extraordinary precision)
\item Ordered rows of seating in parallel aisles between pillars
\item South-to-north slope in the rows of aisles---it has been suggested that the area was washed
\item Paved walkways; upright bricks along aisles (designed for water drainage)
\item Flat, polished stone (61×54×37 cm, dark brown) with clay-pressing wear patterns
\end{itemize}

\textbf{Architectural Logic---Why This Geometry?} A conventional assembly hall, 
religious gathering space, or administrative venue requires only an open floor plan, 
courtyard, or tiered seating, but does not require 20 massive pillars arranged in a 
rigid 4×5 grid. Such a grid imposes geometric order on human movement and partitions 
space into parallel aisles---a design suited to repeated, simultaneous, independent 
operations rather than collective gathering. The pillars themselves are not merely 
structural necessity; they are logistics infrastructure. This geometry solves a specific 
operational problem: enabling multiple independent trading pairs to reconcile accounts 
in parallel, in orderly stations, without interfering with one another. A parliament 
or temple does not require this. AP2PBS Phase 5 reconciliation logic suggests that
obligations were netted out in these parallel trading stations and tablets and
sealings were systematically destroyed and later washed away.

\textbf{Function (Phase 5 reconciliation):}
\begin{itemize}
\item Two counterparties + ULC agents sit in ordered rows with Tablets A/B and sealings
\item Match claim tablets with package sealings; document discrepancies
\item Destroy settled pairs (crushed tablets, broken sealings)
\item Wash debris away via drainage bricks
\item Press new Phase 1 tablets on polished stone; exchange vouchers
\item Each rows of aisles accommodated a many such trading pair---parallel operations at scale
\end{itemize}

\textbf{The polished stone as evidence:} The repeated wear pattern on the polished 
stone surface indicates sustained, regular pressing activity. Within the AP2PBS 
framework, this is consistent with decades of clay-tablet pressing during Phase 5 
reconciliation cycles. Mackay's alternative (leather-working) remains possible on 
physical evidence alone, but the wear pattern sits less naturally with the hall's 
other diagnostic features.

\subsubsection{The Great Hall: Market Intelligence Hub}

\textbf{Scale \& location:} 230 feet × 78 feet; massive open space with multiple
rooms. Wheeler's granary hypothesis rejected by Kenoyer's re-excavation (found
zero grain). Architecture rules out residential, temple, or storage use.

\textbf{Function (seasonal conference):}
\begin{itemize}
\item \textbf{New product showcase:} Merchants introduce goods; agree on new
seal templates and product codes
\item \textbf{Market intelligence exchange:} Prices, harvests, demand signals
        from distant hubs (Magan, Dilmun, regional nodes)
\item \textbf{Coordination among partners:} Periodic gathering of trading
        partners, plausibly supporting relationship maintenance across
        cycles.
\item \textbf{Inclusive participation:} Smaller guilds participate alongside
        hub merchants; share technologies and market intelligence
\end{itemize}

Though this function is inferential rather than directly evidenced.

The scale and layout of the Great Hall---ruled out for residential, temple,
or storage use---are consistent with a venue for periodic large-scale
commercial gathering.

\subsubsection{The Great Bath: Essential Hygiene Infrastructure}

\textbf{Physical infrastructure:} Watertight tank (12m × 7m × 2.4m deep);
bitumen-sealed; colonnaded corridors; elevated, set apart. Industrial-scale,
not luxury.

\textbf{Operational rationale:} Long-distance transit left arriving merchants,
cart drivers, and warehouse handlers covered in salt, silt, and dust---an
unavoidable consequence of maritime, riverine, and overland cartage. The need
for bathing, however, did not depend solely on periodic settlement conferences.
The daily operation of a major commercial center---receiving raw materials,
processing goods, loading and unloading consignments, handling animals and
carts, and dispatching outbound shipments---would itself have created a
continuous occupational hygiene requirement. In the hot climate of the Bronze
Age, without modern ventilation, cooling, or sanitation infrastructure, workers
engaged in these activities would have needed regular access to water for
washing and recovery from dust, salt, sweat, and industrial residue.

Before Phase 5 reconciliation, in which counterparties sat face-to-face for
extended negotiation and physical exchange of tokens, bathing would also have
served a social and presentational function. The Great Bath can therefore be
understood as practical civic infrastructure supporting the daily logistics of
the commercial center, with periodic meetings providing an additional reason
for its use rather than its sole purpose. Within the AP2PBS reconstruction, it
was not merely a luxury or ritual amenity but part of the hygiene infrastructure
required by sustained Bronze Age production, procurement, and exchange.

\textbf{Dual function (proposed):}
\begin{itemize}
\item \textbf{Possible social function:} A shared facility of this scale may
        also have served to mark the transition from transit to commercial
        engagement, though this remains inferential
\item \textbf{Hygiene:} Removal of salt, silt, and dust accumulated in transit,
        consistent with the scale and industrial construction of the tank
\end{itemize}

The Harappan emphasis on bathing infrastructure---individual house baths,
extensive drainage, and this monumental public facility---is more parsimoniously
explained as civic infrastructure supporting commerce at scale than as a
religious or purely domestic amenity.

\subsection{The Integrated Commercial Precinct: Seasonal Cycle}

Three structures formed a single operational system:

\begin{table}[H]
\centering
\caption{Mohenjo-daro commercial precinct: AP2PBS operational roles.}
\begin{tabular}{@{}ll@{}}
\toprule
\textbf{Venue} & \textbf{Role} \\
\midrule
Great Bath & Hygiene + (possible social function, inferential + trust renewal) \\
Great Hall & Market intelligence + periodic partner coordination \\
Pillared Hall & Phase 5 settlement: matching, netting, destroying, signing \\
Polished Stone & New contract pressing \\
\bottomrule
\end{tabular}
\end{table}
\textbf{Seasonal merchant/workmen journey:}
\begin{enumerate}
\item Arrive filthy (cart/boat/maritime transit)
\item Bath: cleanse, mingle with returning merchants
\item College Building: rest and lodging (function proposed; Kenoyer notes
      this structure's original use remains uncertain, dropping the
      unevidenced "priestly" attribution assigned by early excavators)
\item Great Hall: exchange market intelligence, ideas, planning, discuss new products, etc.
\item Pillared Hall: settle past-cycle accounts, match tablets \& sealings, destroy reconciled records
\item Polished stone: press new Phase 1 tablets, exchange vouchers
\item Bath: final cleansing before departure
\end{enumerate}

Year after year, decade after decade. This cycle was the proposed operational
backbone of a peer-to-peer network spanning a million square kilometres---with
no evidence of kings, priests, or central bureaucracy.

\subsection{Logistics Hubs: Waterfront Warehouses}

Beyond the Mohenjo-daro precinct, AP2PBS required staging infrastructure at
nodes throughout the network.  The so-called ``granary'' or ``warehouse'' 
structures at Harappa, Lothal, and other hub sites show no evidence of agricultural
storage; read functionally, they are consistent with inbound and outbound logistics
warehouses--- the physical instantiation of Phase 3 network routing. Their
architecture, when read functionally, reveals a complete workflow for goods reception,
verification, staging, and dispatch.

\subsubsection{Harappa's Mound F: The Inbound and Outbound Logistics Warehouse}

Once identified as a granary, the Mound F structure's function remains uncertain:
excavations have found no evidence that grain or other foodstuffs were stored
there. \citet{marshall1931} describes a large brick complex with sleeper walls
and ventilated rooms, on the northern edge of the mound beside the dry bed of the
Ravi River, linking it geographically to riverine transport and Harappa's
hinterland. The assemblage includes approximately 173 square seals, 36 rectangular
seals, 164 bas-relief tablets, and 215 inscribed tablets, but no comparable
accumulation of sealings \citep{fuls2023catalog}. This combination is consistent
with a facility where durable identity tools and tablets were handled or stored
while temporary package sealings moved on or were removed. It supports a logistics
interpretation, but does not establish the building's function or the proposed
workflow.

\subsubsection{Lothal: Warehouse as Sealing Archive}

\citet{frenez2024} documents 93 sealings in Lothal's warehouse context, about
70 concentrated within one square metre in Block C, Layers 2--3; some bear
multiple seal impressions. This concentration may represent packages temporarily
staged together before handling or dispatch, and an accidental fire preserved
the sealings. In contrast with Mound F's few sealings, Lothal may capture a
moment when temporary package records had accumulated rather than been removed.
The difference is consistent with distinct stages of activity, but does not by
itself demonstrate a shared AP2PBS procedure.

\subsubsection{Kotada Bhadli: Hinterland Caravanserai and Distributed Way-Stations}

\citet{shirvalkar2025caravanserai} identifies features at Kotada Bhadli in
Kachchh consistent with caravanserai infrastructure, while evidence of dairy
production may indicate provisioning for people or animals. The site could have
supported rest and movement along overland routes connecting larger centres,
complementing riverine and maritime hubs such as Mound F and Lothal. Its role as
a transfer point within the proposed network remains an interpretation, not a
directly documented function.

\section{Discussion}

\subsection{Network Effects and Emergent Standardization: From 
Commonwealth to Coordinated Scale}

The paradox posed in the Introduction is how the Harappan Civilization achieved
extraordinary standardization (USS, calibrated weights, gridiron cities, uniform
brick, coordinated drainage) without centralized political authority. We propose
that network effects operating on independent merchant enterprises resolve it.

The statistical evidence from Sections~\ref{sec:unicorn_stats}
and~\ref{sec:site_stats} converges on three observations. First, the Unicorn Seal
System is a shared authentication convention across 19 sites and 139 offering
stand types, each stand functioning as a merchant guild or trading entity. Second,
the diachronic trajectory (Section~\ref{subsubsec:diachronic_scurve}) is
consistent with an S-curve of sparse emergence, rapid expansion and slower growth.
Third, the power-law distribution of offering stands
(Section~\ref{subsubsec:offering_stands}) is consistent with preferential
attachment, as new traders copy a successful commercial convention.

These patterns are what network externalities would produce. When independent
merchants converge on shared protocols because adoption becomes individually
rational, standardization emerges from the bottom up, with an S-curve adoption
trajectory and a power-law frequency distribution. No central authority need have
issued orders: merchants adopted the convention because it widened their trading
opportunities and counterparty confidence, and once critical mass was reached each
adopter reinforced the standard through positive feedback. The result is
coordination without a coordinator, akin to railways, interstate highways,
telecommunications and the internet, where standardization is likewise diagnostic
of network externalities. The hubs of such a network, the multi-site stands and
the large sites, can be read as the ``heterarchies'' that
\citet{vidale2018heterarchic} describes as ``the power systems that ruled the
polycentric'' cities of the Harappan Civilization: several centers of influence
without a single apex. The institutional requirements of this self-organization
are detailed in the next section.

\subsection{Seal System and associated AP2PBS produces scale-free signatures}

All the distributions---seal motifs, guild sizes and site seal counts---exhibit
the heavy-tailed, hub-dominated structure characteristic of scale-free networks.
The evidence rests on multiple pillars:

\begin{enumerate}
    \item \textbf{Seal motifs follow a power law:} the Unicorn dominates the
    iconographic repertoire and the other motif classes fall off rapidly; the
    exponential is rejected, while lognormal and truncated power law cannot be
    distinguished from the power law in this small sample.
    \item \textbf{Guild sizes follow a power law:} a few dominant guilds account for
    a disproportionate share of seals, while many guilds appear only once.
    \item \textbf{The power law persists across periods:} guild-size distributions
    remain heavy-tailed from 3A to 3C, so the hub-and-spoke structure is not an
    artifact of a single period, although the early samples are small.
    \item \textbf{Stand and attributes form a shared convention:} stand type is
    strongly associated with the other attributes, and the same stand tends to carry
    the same attribute values across sites, which is consistent with a coordinated
    standard.
    \item \textbf{Site seal counts follow a power law:} a small number of mega-hub
    sites dominate, although excavation history contributes to this.
    \item \textbf{Unicorn adoption is consistent with an S-curve:} a lag, rapid
    expansion and slower growth, suggestive of network effects and standard
    diffusion.
    \item \textbf{Top guilds and singletons both span sites:} top-ranked guilds
    appear across most sites, consistent with high connectivity, and singleton guilds
    also appear at multiple sites, which is possible where common infrastructure
    exists.
\end{enumerate}

This power-law structure---market hierarchy encoded in offering stand variants
---explains how asynchronous peer-to-peer trade could operate at scale. Hub
guilds (few in number, high in volume) could coordinate across the network.
Peripheral guilds (many in number, low in volume) could participate through hub
intermediation. The scale-free architecture made AP2PBS operationally feasible:
a merchant from a peripheral guild could trade with strangers anywhere in the
network by leveraging hub guild authentication and logistics infrastructure.

\subsection{The Unicorn Seal as Multi-Layer Operational Contract}

The unicorn seal was not a decorative symbol or administrative marker.
It was a \textbf{multi-layer operational contract} encoding the full commercial
transaction lifecycle:

\begin{table}[H]
\centering
\caption{Unicorn seal functions across AP2PBS workflow phases - both directions.}
\begin{tabular}{@{}p{3.5cm}p{5cm}p{5cm}@{}}
\toprule
\textbf{Form} & \textbf{Function} & \textbf{AP2PBS Role} \\
\midrule
Seal on steatite (permanent) & Long running contract \& guild identity & Phase
        1 negotiation; Phase 1 sealing; reusable across cycles \\
Sealing A/B (impressed on package) & Integrity proof; tamper-evident; executed
        delivery & Phase 2 and 4 shipment; proves origin, integrity and sealing receipt \\
Tablet A/B with flat back & Claim voucher + discrepancy record + routing slip
        & Phase 3 routing; Phase 4 claim voucher; Phase 5 reconciliation; bilateral marking of shortages \\
\bottomrule
\end{tabular}
\end{table}

Each seal impression is both \textbf{unique} (specific transaction) and
\textbf{replicable} (same seal carved by same carver maintains visual continuity).
This dual property allowed:

\begin{itemize}
\item \textbf{Authentication across time:} A seal carved once could be reused for
        decades, allowing merchants separated by years to verify their
        counterparty's identity
\item \textbf{Confirmation across distance:} The confirmation bundle (unicorn + offering
        stand + attributes) could be recognized at any waypoint without language
        intermediation
\item \textbf{Legal continuity:} Jamison's multi-site seal groups (21 of 55 groups
        span multiple sites, 38\% cross-site distribution) represent guild networks
        persisting across decades and geography
\end{itemize}

The offering stand variants encode market structure directly into the seal iconography.

\subsection{Scale-Out: Trade Logistics and the Gridiron across the Harappan Civilization}

At the Early to Mature transition, it seems like Harappan urbanism just scaled out.
The merchants and traders were not trying to scale out---it just looks that way in
the archaeological record. They seem to have been singularly focused on just trying
to remove any friction that impedes trade.

Possehl talks about Indus ideology, invoking European revolutionary era nihilism as
the driving ideology. His assertion seems to be grounded in two particular facts: one,
that there seem to have been conflagrations in certain sets of sites, and the fact that
``there is a proclivity for Indus sites to be founded on virgin soil; 755 out of 1058
sites, or 71 percent. Of the 523 Early Harappan sites we know of, 324 (62 percent)
were abandoned prior to Mature Harappan'' \citep{possehl2002indus}.

We posit that the scaling of trade across the Harappan region is a parsimonious
explanation. The archaeological record shows toy carts and flat-bottomed boats
increasing as the Early Harappan turns Mature, and the humble ox cart has been
under-appreciated as a driver of the earliest urban gridiron.

The logistics of this network plausibly grew out of pastoralism. Pastoral groups
already moved herds between uplands, plains and river valleys, and so knew the
seasonal routes, the water and fodder stops, and the people at either end. As
trade developed, some of them plausibly became intermediaries specializing in
long-distance movement, first with pack animals and then with carts. Each step
added a skill: managing teams of oxen and maintaining carts, choosing and keeping
routes, and knowing what was scarce and what was in surplus in distant regions.
This last skill, information arbitrage, made them more than carriers. An
intermediary who saw both ends of a route could coordinate shipments across the
network, lower transaction costs, and let strangers trade with confidence. These
are the operators the seal system and the AP2PBS protocol would have served.

Cart geometry---axle width, turning radius, load capacity---set requirements that
the organic, winding lanes built for foot and donkey traffic could not meet. Main
streets had to let two paired-ox carts pass each other beside pedestrians, and the
rounded building corners found at some sites are consistent with protecting walls
from passing carts. The gridiron was the capillary system of the cart network,
feeding the arteries of the Sindhu and the sea: straight streets reduced friction
and perpendicular intersections simplified routing. Carts may thus have accelerated
gridiron urbanism, in addition to the light advantages of a north--south alignment.

We propose that the gridiron was laid down when trade scaled, and that it came
first as a cart track outside the regular settlement where carts lined up to load
and unload, and not as a town plan. Carts needed a broad, straight track
just outside the older settlement, where loading, marshalling and storage were
convenient. Merchants who sited themselves along the track gained access to the
trade, and competition drew others to it, so that within perhaps a generation the
settlement's centre of gravity moved to the gridiron. The old winding settlement
was then left, burned or built over.  Possehl noted this pattern, and there is
emerging evidence for this pattern. This is the dynamic we offer against Possehl's
``nihilism'': the abandonment of many Early Harappan sites and the founding of new
ones on virgin soil need not reflect an ideology of rupture, since a shift of the
settlement to its new trade axis would produce the same record. Perhaps the citadels
are sites that were built over later. The lower gridiron-layout town is where wealth
in a barter economy became concentrated, and urbanism seems to have followed the
gradient of increasing trade.

The Early-to-Mature Harappan transition looks in the archaeological record like a
phase transition---not gradual linear growth but a relatively rapid nonlinear jump
to a new organizational level. That's exactly what growth and preferential attachment
network dynamics produce. Once a few nodes adopt the fast (for the Bronze age)
logistics-optimized model successfully and begin attracting disproportionate trade
traffic, the competitive pressure on remaining non-optimized sites becomes intense
and the cascade is fast.  The Mature-to-Late Harappan looks like the reverse phase
transition---something hit the hubs or top guilds that administered ULC very hard.
Late Harappan seems to go back to geometric seal motifs, with local, perhaps
family-based trade continuing. Perhaps the local, family-based trade always existed
through the Mature Harappan alongside the high-throughput trade across vast distances.

The Mature Harappan package---grid cities, standardized weights, seals, carts,
flat-bottomed and maritime boats---isn't a collection of independent innovations
that accelerated trade. It's a single integrated response to trade network scaling,
appearing together because the competitive pressure from trade scaling that drove
one drove all of them simultaneously.

This mirrors modern infrastructure revolutions (arrival of railroads - reorganized
cities, inter-state highways - the suburban sprawl, internet - information
networks): slow build-up, then rapid reorganization into a new scaling regime,
followed by eventual transformation when conditions change.

The absence of palaces, war, and religion, the ubiquitous Unicorn, which
we have demonstrated to be aligned with commercial and trade networks, provides
evidence fitting a decentralized, economically driven complex commercial network
rather than a command hierarchy.

We position Harappan urbanism as a classic example of how complex societies can emerge,
thrive, and transform without rulers or priests. The driver of the Mature Harappan
was the rapid scaling of the trade network and a civilizational outlook oriented to
the reduction of any impediments to trade.

\subsection{Periodic Assembly as Standardization Engine}

The remarkable uniformity of Harappan standards---brick ratios, weight stones, street
grids, seal formats---across 1,500+ sites with scarce evidence of central coercion is
explained by the periodic meetings held at the hub sites. The expertise related to
all the features that are diagnostic of Harappan civilization would disseminate
periodically, the Sindhu template would spread as though by osmosis.

This kind of periodic assembly of a variety of peer expertise perhaps is the most
parsimonius explanation of the remarkable standardization over a vast geography.
These meetings were driven by the different cycles of the wide array of products
and commodities that the Harappans traded - including bricks, sanitation and other
technology that is diagnostic of Harappan civilization.

This kind of peer-level periodic meetings was unprecedented historically, perhaps.
Never before were there such periodic assembly of expertise at a central places in
civilizational level histories during the third millennium BCE.

Voluntary adoption spread because:

\begin{itemize}
\item \textbf{Demonstration effect:} Hubs like Mohenjo-daro showed operational
        efficiency of standardization
\item \textbf{Economic advantage:} Merchants preferentially traded at standardized
        sites; peripheral sites faced competitive disadvantage
\item \textbf{Portable expertise:} Master carvers and engineers traveled between
        sites, disseminating standardization
\item \textbf{Switching costs:} Once a site adopted standardized bricks and weights,
        reverting to non-standard materials meant inventory incompatibility
\end{itemize}

\subsection{Scale-Free Robustness Explains 700-Year Longevity}

Scale-free networks are robust to random failures but vulnerable to targeted hub
disruption \citep{albert2000error}. The Mature Harappan urban phase ($ \sim$700
years) exhibited this robustness:

\begin{itemize}
\item Loss of small peripheral guilds or sites had minimal network impact
\item Merchants could route around failed sites using alternative hub paths
\item The AP2PBS protocol itself had no critical single point of failure---any two
merchants could authenticate transactions using standardized seals and weights
\item Decentralized expert assembly meant no single person or office whose
removal could collapse standardization
\end{itemize}

This explains why the civilization sustained itself across such vast distances
and time scales despite lack of centralized authority.

\subsection{Hub Collapse and Systemic Cascade Failure}

The Late Harappan transition represents network-level cascade failure, not
gradual decline.  When hub sites failed---whether from environmental stress
(Ghaggar-Hakra river shifts), trade disruption (Mesopotamian network collapse),
or other factors---the system unraveled suddenly:

\begin{itemize}
\item \textbf{Loss of standardized weights:} Without hub metrologists maintaining
calibration and distributing reference stones, peripheral merchants lost assurance
of weight consistency. Trading friction returned.
\item \textbf{Loss of seal carving expertise:} Without hub artisans maintaining
standardized seal templates, new seals could not be commissioned. Merchants
reverted to non-standardized, often geometric, seals.
\item \textbf{Loss of infrastructure coordination:} Without periodic expert
assembly, bricks degraded without replacement (architectural standards lost),
drainage systems failed, grid-planned settlements became economically
unmaintainable.
\item \textbf{Urbanism dissolution:} Cities could not sustain themselves without
the trade surplus that hub coordination generated. Urban depopulation was rapid.
\item \textbf{Unicorn seal disappearance:} Hub guild production ceased. Peripheral
guilds, unprepared for independent seal manufacturing, abandoned the standardized
system.  Unicorn seals vanish from the archaeological record essentially overnight
in regional terms.
\end{itemize}

The archaeological signature is sudden, not gradual: vanishing urbanism, loss of
standardized weights, end of brick uniformity, reversion to geometric seals,
fragmentation into local trade networks. This is precisely what scale-free
network collapse predicts when hubs fail.

\subsection{Script Symbols as Operational Notation System}

Decipherment programs, such as the Sangam Rebus theory, Proto-Indo-Aryan readings
and general linguistic approaches, begin by assuming that \textbf{the script
primarily encodes language}, and ask which language the signs record. The present
work begins instead from the trade context in which the Unicorn Seal System (USS)
appears, and asks what problem the seals were solving pertain to trade. We therefore
ask what information a Bronze Age long-distance trade protocol would \textit{need to
communicate}. Merchants separated by hundreds of kilometers who exchange goods
asynchronously must agree beforehand on standard weights and measures, quality
grades, ownership authentication, routing, receiving authority and settlement
rules; without these, delayed exchange cannot be verified. The signs are
therefore read first as logistical categories (commodity, quantity, container,
destination, weight class) and not as words. A boat-like sign is a candidate
for a vessel class because vessel type is information such a system would need,
not because it resembles a boat. The two programs are different research
programs, and each can be judged by how well it explains the archaeological
observations: brevity, repetition and the absence of personal identification.

The script appears on seals, tablets and sealings. Inscriptions are short
(typically 4--5 signs), highly repetitive, and conspicuously lack personal names
and place names. These properties are puzzling if the script records speech, but
are expected of operational notation. \citet{mukhopadhyay2023} concluded that the
script does not encode anthroponyms or toponyms but ``trade and commodity
control,'' and the stand statistics are consistent with her conclusion: identity
was carried visually by the Unicorn, the stand and its attribute bundle, so
the inscription did not need to name the parties. Identical signs recur across
sites, which indicates standardized codes and not local linguistic variation, and
metrology signs occur in spatial association with hexahedral weights, which
suggests a notational link between the script and standardized measures.

The rare signs fit the same reading. Of roughly 400 distinct signs, 113 occur
only once (hapax legomena) and 47 only twice. If a guild introduced a new
commodity, it could agree a provisional sign with its trading partner; the sign
would enter the standard repertoire if trade in the product continued, and would
never recur if the transaction was a one-off or the product was discontinued.
Singleton stands (representing commercial entities) could likewise generate signs
used in isolated transactions.  This is a hypothesis consistent with the data, not
a demonstration, but it accounts for the long tail of rare signs without assuming
incomplete transmission of unknown words.

The three elements of a Unicorn seal---the animal, the stand and the
inscription---were rarely manufactured separately, yet decipherment studies
usually treat the script alone and glyptic studies the iconography alone. Across
more than 3,700 seals the same package recurs, which suggests a unified
institutional technology: the animal supports authentication through visual
recognition, the stand identifies the commercial entity (guild, warehouse or
merchant house), and the script supplies operational metadata for goods in
motion. The package is characteristic of periods 3A and 3B, the height of
standardization and inter-regional exchange. In period 3C script-only tablets
appear, but they are scarce and occur mainly in workshop areas, suggesting
internal production documentation and not authentication tokens for exchange
between parties. The framework can thus be evaluated without identifying
individual signs: it must explain why Unicorn seals dominate, why inscriptions
are short, why sealings are scarcer than seals, why weights are standardized over
so wide an area, and why sign order is so regular. If the institutional model
(AP2PBS coordinated through the ULC) explains these observations better than
competing models, it gains credibility, and what individual signs denote becomes
a secondary hypothesis requiring its own evidence.

\subsection{Etymology that traces its origins in Harappa}

The Harappan civilization was a network of merchants and guilds operating
across vast distances. Their periodic meetings at hub sites were perhaps the
\textbf{commercial sabh\={a}s} where these guilds coordinated trade and their
ubiquitous standards, perhaps even the entire Sindhu template. These were
likely the institutional precursors to the forms that would later be documented
as \textit{Śreṇi}, \textit{Nigama}, \textit{Pūga}, \textit{Gana}, and \textit{Sārtha}---the
vocabulary of merchant associations and market bodies that appears in texts from
the Mauryan and Sangam eras. If this continuity held, the Harappan sabha was where
the earliest guilds and merchant associations were formed.

The \textit{\'{S}re\d{n}i}---a term appearing in Sanskrit and Prakrit texts from the 
Mauryan era onward, likely preserved from pre-Aryan language substrates of South 
Asia---denotes both ``row'' or ``line'' and, by extension, ``merchant guild'' or 
``professional association.'' The Pillared Hall materializes this concept into 
operational reality. Traders settled their obligations literally sitting in 
rows---the \textit{\'{S}re\d{n}i} as both spatial arrangement and corporate body.
Even prior to Mature Harappan times, perhaps two row of sealings were tallied by
counterparties to net a barter settlement when they met for periodic reconciliation.

The figure administering this ordered settlement was the \textit{\'{S}re\d{s}\d{t}hin}, 
whose title traces to this coordinative function. The word's preservation across 
centuries and language families suggests institutional continuity: the commercial 
practices that generated the term's meaning in the Harappan era persisted, in 
recognizable form, into the documented guild systems of classical India.

\section*{Acknowledgments}

This work is deeply indebted to the meticulous typological scholarship of
Gregg M.~Jamison, whose doctoral dissertation \citep{jamison2017} provided the
comprehensive data on unicorn seal stylistic variation that made this analysis
possible. Jamison's painstaking cataloguing of 500 seals across eleven attributes
and nineteen archaeological sites represents years of careful archaeological work.
While we reinterpret his data through a different analytical lens, we wish to
emphasize that the empirical foundation is entirely his. Any errors in
interpretation are our own.

Statistical analysis, code development, and manuscript preparation were assisted by
GitHub Copilot (Claude, Anthropic). Various AI chatbots were used including
Claude for building the arguments and ChatGPT (adverserial reviews) and Gemini
(cooperative review) for reviewing the paper draft. Gemini, Mistral Vibe and
Deepseek were used for, for literature research and review of operational details
of AP2PBS.

\section*{Data Availability}

All code, data, and reproducibility documentation are openly available in the
GitHub repository:
\href{https://github.com/dmahesh2869/harappan-seal-analysis}{dmahesh2869/harappan-seal-analysis}.

The repository is organized in three layers:
\begin{itemize}
        \item \textbf{Core reproducibility pipeline:} primary scripts, validation checks,
        and run instructions needed to regenerate the manuscript's main statistical
        results, tables, and figures.
        \item \textbf{Supplementary analyses:} additional scripts used for robustness
        checks and exploratory diagnostics (e.g., stand hierarchy, bundle consistency,
        singleton dispersion, and diachronic analyses).
        \item \textbf{Versioned outputs:} generated CSV, TXT, and PNG artifacts used to
        audit and verify claims in the manuscript.
\end{itemize}

All analyses are deterministic and reproducible in Python 3.8+ with
\texttt{powerlaw} (v2.0.0).

\section{Conclusion}

This paper has argued that the Harappan Civilization is best understood as a merchant
commonwealth. It was a merchant commonwealth organized as a long distance barter trade
network---unprecedented in human history and never fully replicated since.

At its operational core lay the what we propose as periodic meetings, the USS,
the AP2PBS (Asynchronous Peer-to-Peer Barter Settlement) protocol or something akin
to it and the Unicorn Logistics Confederation (ULC)---a confederation of leading merchant
guilds---provided the institutional framework. It guaranteed weight standardization,
authorized seals, arbitrated disputes, secured delivery, underwrote counterparty
risk, and coordinated multimodal logistics. Six functions that made the system work.

This civilization demonstrates how complex societies emerge, thrive, and transform
without script (in the linguistic sense), without rulers, and without priests. The
driving motive was perhaps simple: reduction of friction to trade. Every innovation
served that motive. The USS was iteratively refined to reduce friction and enhance
the efficiency of trade. Every architectural form followed that function.

It is poetic that when Louis Sullivan articulated his architectural dictum at
the turn of the twentieth century, he was ushering in the age of vertical
urbanism---soaring skyscrapers rising toward the sky in response to the
density requirements of machine-age cities. Yet the Harappans, more than
4,600 years earlier, had perfected Sullivan's very principle when they
organized the horizontal gridiron. Every alley, every wall thickness,
every drain, every warehouse orientation followed directly from the
functional requirements of long-distance cartage and exchange. They
proved that function could be sovereign without abandoning beauty---that
the most elegant city might be the most economically efficient one.

\vspace{0.3in}

\textit{Form ever follows function. This is the law.} --- Louis H. Sullivan

\end{document}